\documentclass[journal=jctcce,manuscript=article,layout=onecolumn]{achemso}

\usepackage[version=3]{mhchem}
\usepackage{chemformula}
\usepackage{subcaption}
\usepackage{booktabs}
\usepackage{tabularray}
\usepackage{multirow}
\usepackage{array}
\usepackage{enumitem}
\usepackage[sectionbib]{chapterbib}

\newcommand*\angstrom{\textup{~\AA}}

\author{Joe Gilkes}
\affiliation[Chem]{Department of Chemistry, University of Warwick, Gibbet Hill Road, CV4 7AL Coventry, UK}
\alsoaffiliation[HetSys]{EPSRC HetSys Centre for Doctoral Training, University of Warwick, Gibbet Hill Rd, CV4 7AL, Coventry, UK}
\author{Mark Storr}
\affiliation[AWE]{AWE Plc, Aldermaston, UK}
\author{Reinhard J. Maurer}
\affiliation[Chem]{Department of Chemistry, University of Warwick, Gibbet Hill Road, CV4 7AL Coventry, UK}
\alsoaffiliation[Phys]{Department of Physics, University of Warwick, Gibbet Hill Road, CV4 7AL Coventry, UK}
\email{r.maurer@warwick.ac.uk}
\author{Scott Habershon}
\email{s.habershon@warwick.ac.uk}
\affiliation[Chem]{Department of Chemistry, University of Warwick, Gibbet Hill Road, CV4 7AL Coventry, UK}

\title[Understanding and improving transferability in machine-learned activation energy predictors]
{
\singlespace Understanding and improving transferability in machine-learned activation energy predictors
}

\begin{document}

\begin{center}
    UK Ministry of Defence \copyright \ Crown Owned Copyright 2025/AWE
\end{center}

\begin{tocentry}
    \includegraphics[]{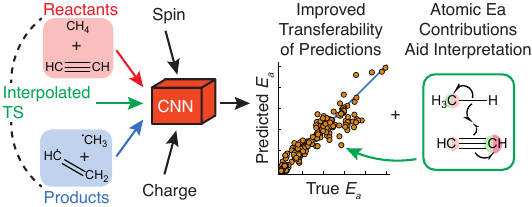}
\end{tocentry}

\begin{abstract}
    \singlespace
    \noindent The calculation of reactive properties is a challenging task in chemical reaction discovery. Machine learning (ML) methods play an important role in accelerating electronic structure predictions of activation energies and reaction enthalpies, and are a crucial ingredient to enable large-scale automated reaction network discovery with $>10^3$ reactions. Unfortunately, the predictive accuracy of existing ML models does not yet reach the required accuracy across the space of possible chemical reactions to enable subsequent kinetic simulations that even qualitatively agree with experimental kinetics. Here, we comprehensively assess the underlying reasons for prediction failures within a selection of machine-learned models of reactivity. Models based on difference fingerprints between reactant and product structures lack transferability despite providing good in-distribution predictions. This results in a significant loss of information about the context and mechanism of chemical reactions. We propose a convolutional ML model that uses atom-centered quantum-chemical descriptors and approximate transition state information. Inclusion of the latter improves transferability for out-of-distribution benchmark reactions, making greater use of the limited chemical reaction space spanned by the training data. The model further delivers atom-level contributions to activation energies and reaction enthalpies that provide a useful interpretational tool for rationalizing reactivity. 
\end{abstract}

\newpage
\section{Introduction}

Complex chemical reaction networks (CRNs) arise across a wide variety of fields, from catalysis and combustion modelling to atmospheric chemistry and biological synthesis.\cite{Gopalkrishnan2011,Margraf2023,Xu2011,Chen2023,Rimmer2016,Liu2020,Loskot2019} A wide variety of automated reaction discovery (ARD) schemes are available, including contact-map-based sampling strategies,\cite{Zimmerman2013,Suleimanov2015,Habershon2015,Habershon2016,Gilkes2024} heuristic rule-driven methods,\cite{Susnow1997,Gao2016,Rappoport2014,Bergeler2015} and strategies that manipulate the potential energy surface (PES) to drive the exploration of chemical reaction-space.\cite{Ohno2004,Maeda2009,Maeda2016,Wang2014} Such ARD methods have been reviewed elsewhere, but a common thread is that --- inevitably --- making predictions about the experimentally-observable properties of CRNs demands evaluation of the properties of individual elementary reactions, primarily reaction enthalpy $\Delta H_r$ and activation energy $E_a$.\cite{Coley2020,Dewyer2018}

Given the scale of possible ARD-generated CRNs --- with some recent examples comprising $10^{3} - 10^{4}$ possible elementary reactions --- direct \emph{ab initio} evaluation of reaction properties is time-consuming and resource intensive. As such, a variety of different machine learning (ML) strategies have been employed to predict activation energies and/or reaction enthalpies after training on suitable datasets, typically calculated at the level of density functional theory (DFT). Recent examples include the Chemprop neural network (NN) model,\cite{Yang2019,Grambow2020Chemprop} as well as the KPM model.\cite{Ismail2022} These models typically encode reaction information using `difference' vectors --- in other words, subtracting a reactant descriptor from a product descriptor, resulting in a descriptor representing the important functional group changes during the chemical reaction that may subsequently be used for ML. In applications to a dataset of $>10^{4}$ organic chemical reactions reported by Grambow \textit{et al.}, \cite{Grambow2020} such models provide predictive accuracy in activation energies to within $5$ kcal/mol, which represents the current state-of-the-art.

In recent work, we combined our KPM activation-energy predictor with a streamlined ARD strategy that iteratively constructs a CRN using kinetic simulations with ML-based rate constants, derived from predicted activation energies,  to guide the exploration of relevant regions of chemical reaction space.\cite{Gilkes2024} Our results highlighted the challenge of accurately modelling complex CRNs with ML-based rate predictions, with some final product concentrations disagreeing significantly with previous shock-tube experiments. While KPM shows sufficiently low test set errors within training, deeper analysis performed here reveals a severe lack of transferability and generalisability to reactions outside the distribution of training reactions. This is echoed in similar contemporary ML strategies.

The goal of this article is to explore these failures in more depth to pinpoint key transferability problems with current structure-based ML strategies like KPM and Chemprop and to identify future directions; as we show below, analysis of the descriptors used in these models (and their resulting view of chemical reaction space) explain their lack of transferability to seemingly similar reactions. Using these new insights, we propose a new NN architecture using atom-centered descriptors obtained from quantum-chemical calculations in combination with approximate transition state geometries to improve transferability; the results of this strategy are encouraging, albeit with further room for improvement. 

The remainder of this article is organized as follows. First, in Section \ref{sec::benchmark}, we perform a thorough benchmarking of contemporary ML-based strategies for reaction-property prediction (KPM, Chemprop and NeuralNEB), highlighting their disparate views of chemical reaction space. In Section \ref{sec::spahm}, we introduce a novel NN architecture that offers a much better route to reaction-property prediction with improved transferability. Finally, we show how such atom-based ML models offer new chemical insights, providing an easily explainable view of reactivity for organic molecules.

\section{Benchmarking Contemporary Models}\label{sec::benchmark}

Before exploring new ML models with improved transferability to previously-unseen reactions, it is vital to understand both how and why KPM and similar ML models fail at the task of predicting reactive quantities in some areas of chemical reaction space. It is also important to quantify how different these regions must be from the reactions of the training set of a given model in order to yield poor predictions.

For this reason, an additional validation dataset of reactions were first generated from the CRN produced at the end of our previous work.\cite{Gilkes2024}. This CRN, generated with the Kinetica.jl package and labelled $C_4^{I, 0.01}$, explores the chemical reactions that may occur during ethane pyrolysis at 1000 K. It therefore contains over 8,000 reactions of pure hydrocarbons with no heteroatoms, many of which feature open-shell free radical species that can be formed at such high temperatures. Despite the apparent simplicity of such reactions, rate constants based on KPM's $E_a$ predictions were not accurate enough to obtain even qualitatively correct CRN kinetics when compared to experimental data. While we postulated that this was likely due to a missing entropic contribution to the rate constants used for guided exploration of chemical reaction space, the accuracy of KPM's $E_a$ predictions on these ARD-generated reactions has not yet been systematically tested.

To examine this potential source of error, minimum energy paths (MEPs) and transition state (TS) geometries of these generated reactions were isolated using the climbing image nudged elastic band (CI-NEB) method, as detailed in Fig. S1 of the \textit{Supporting Information}. Calculations were performed in the NWChem electronic structure code at the same level of theory as the original datasets of Grambow \latin{et al.} on which KPM was trained (namely DFT with $\omega$B97X-D3 hybrid functional and def2-TZVP basis set).\cite{NWChem,Grambow2020,Grambow2020_raddata} This ensures that the resulting DFT-calculated activation energies are as comparable to the activation energy predictions of KPM as possible. TSs were confirmed through vibrational analysis, by checking that each had only one imaginary mode present.

Of the 637 total reactions from previously-generated CRN $C_4^{I, 0.01}$ that were put through this workflow, 381 produced converged TS geometries that could be used to calculate accurate activation energies. A selection of the converged reactions are presented in Fig. S2. While some of these reactions were barrierless or comprised a single energetic barrier, others consisted of multiple energy barriers. The multi-step reactions were a result of the connectivity-based reaction exploration algorithm used to generate the initial CRN, which is less restrictive to the types of reactions explored than algorithms based on, for example, bond-order matrices. While these multi-step reactions can in principle be further reduced to multiple single-step reactions by iteratively performing further NEB calculations, we avoid this complication here by simply removing these reaction types. Ultimately, this leaves a validation set of 147 converged, single-step reactions from network $C_4^{I, 0.01}$ (extended to 294 reactions by also extracting the reverse of each calculated `forward' reaction) with which ML models could be benchmarked for potential out-of-distribution accuracy.

As shown below, this allows quantification of the error incurred by generalising ML $E_a$ predictions to unknown regions of chemical reaction space, while simultaneously enabling visualization of the space occupied by both the original training set and this new validation set.

\subsection{KPM}

To test transferability, a KPM model was used to predict the activation energies of the new validation set of chemical reactions.\cite{Ismail2022} This model was the same as was used in our previous work, and had been trained on a combination of two datasets by Grambow \latin{et al.}: a general purpose dataset of organic reactions comprised of C, H, O and N atoms in charge-neutral and ionic reactions, and a secondary dataset of radical reactions of the same elements which were excluded from the primary dataset.\cite{Grambow2020,Grambow2020_raddata} Both datasets contained zero-point energy-corrected activation energies and reaction enthalpies for all reactions within. It was expected that inclusion of the secondary dataset within the training data would extend the model's understanding of chemical reaction space, allowing the validation set reactions to be predicted from within the training distribution. Training on this combined dataset did introduce some additional error into the test set predictions compared to the original KPM model of Ismail \latin{et al.} (MAE = 1.98 kcal/mol, RMSE = 5.17 kcal/mol), although not to an extent that the authors' original target of 2--6 kcal/mol was exceeded.\cite{Ismail2022}

\begin{figure}[h!]
    \centering
    \includegraphics{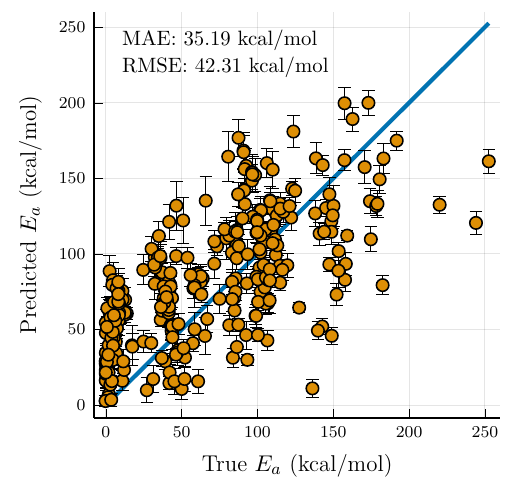}
    \caption{Correlation plot for KPM $E_a$ predictions of $C_4^{I, 0.01}$ validation set.}
    \label{fig:ml-kpm-ood-test_only}
\end{figure}

This model was applied to the reactions of the validation set and their predicted activation energies were compared against their corresponding DFT results. All DFT calculations were corrected using the zero-point energies of their respective geometries, to match the energies provided within KPM's training data. The correlation plot for these results is shown in Fig. \ref{fig:ml-kpm-ood-test_only}.

The poor agreement between the DFT and KPM-predicted activation energies in Fig. \ref{fig:ml-kpm-ood-test_only} helps to explain why our previous kinetic simulations struggled to qualitatively reproduce the experimental reference data. While there is a general correlation between the two sets of activation energies, the errors across the board are simply too large to yield accurate kinetics results - especially when one considers the exponential relationship between $E_a$ and reaction rate constant in the Arrhenius equation, which causes these large prediction errors to be exacerbated.

We initially postulate that the large errors in the validation dataset arise due to the lack of representation of the validation set reactions in the KPM training set (noting that the secondary dataset of reactions of free radical species is one tenth of the size of the primary dataset of ionic and charge-neutral reactions). Without enough similar reactions to learn from, it is sensible to assume that KPM struggles to generalise to new areas of chemical reaction space. To test this hypothesis, we added half of the reactions in the $C_4^{I, 0.01}$ validation set to the training dataset and trained a new KPM model. The held back remainder of the validation set was applied to this model, as we hoped to see \emph{some} improvement in their prediction accuracy when KPM had further prior knowledge of the relevant chemical reaction space. This resulted in marginally improved validation set errors (MAE = 31.07 kcal/mol, RMSE = 40.13 kcal/mol; correlation plots are shown in Fig. S3), but this could easily be the result of averaging over a smaller validation set.

 It therefore appears that KPM fails to generalise to the reactions in network $C_4^{I, 0.01}$, even when given similar reactions to learn from. However, this conclusion was complicated by the ensemble uncertainties across the predictions of this retrained model --- the uncertanities for the out-of-distribution validation set reactions were just as small as for the reactions in the new training set. If the reactions of the validation set were truly out-of-distribution and KPM was failing to generalise to the areas of chemical reaction space they represent, it would be expected that the individual NN ensemble members would disagree over their predictions, yielding a higher predictive uncertainty. The fact that the opposite trend is observed suggests that the large prediction errors in the validation set cannot be explained through poor training alone: a deeper level of analysis is required.

\subsubsection{Chemical Reaction Space Analysis}

To begin this further analysis, we first sought to understand whether the reactions of the validation set truly laid outside of the distribution of reactions in the KPM training set. To assess this, the reaction descriptors used in KPM were recalculated for both the training set and for the $C_4^{I, 0.01}$ validation set. These descriptors are formed from the element-wise difference in 1024-bit Morgan fingerprints\cite{Morgan1965,Zhong2023} between reactants and products for each reaction, concatenated with the zero point energy-corrected enthalpy change of that reaction to create a 1025-length descriptor vector.

Initially, we calculated Tanimoto coefficients for every pair of reaction descriptors across both training and validation datasets to measure their similarity.\cite{Tanimoto1958,Jaccard1912} We compare the similarities between the reactions of the training set and the reactions of the validation set to the self-similarity between all pairs of reactions in the training set in Fig. S4. While there are some differences, this form of analysis only reveals that the validation set is as similar to the training set as the training set is to itself, suggesting that the validation set should lie within the training distribution. However, Tanimoto similarity is limited to directly comparing features across two descriptors without accounting for the correlation between features, so this may not reveal the true differences between the reaction sets.

\begin{figure}[h!]
    \centering
    \includegraphics[width=240.71031pt]{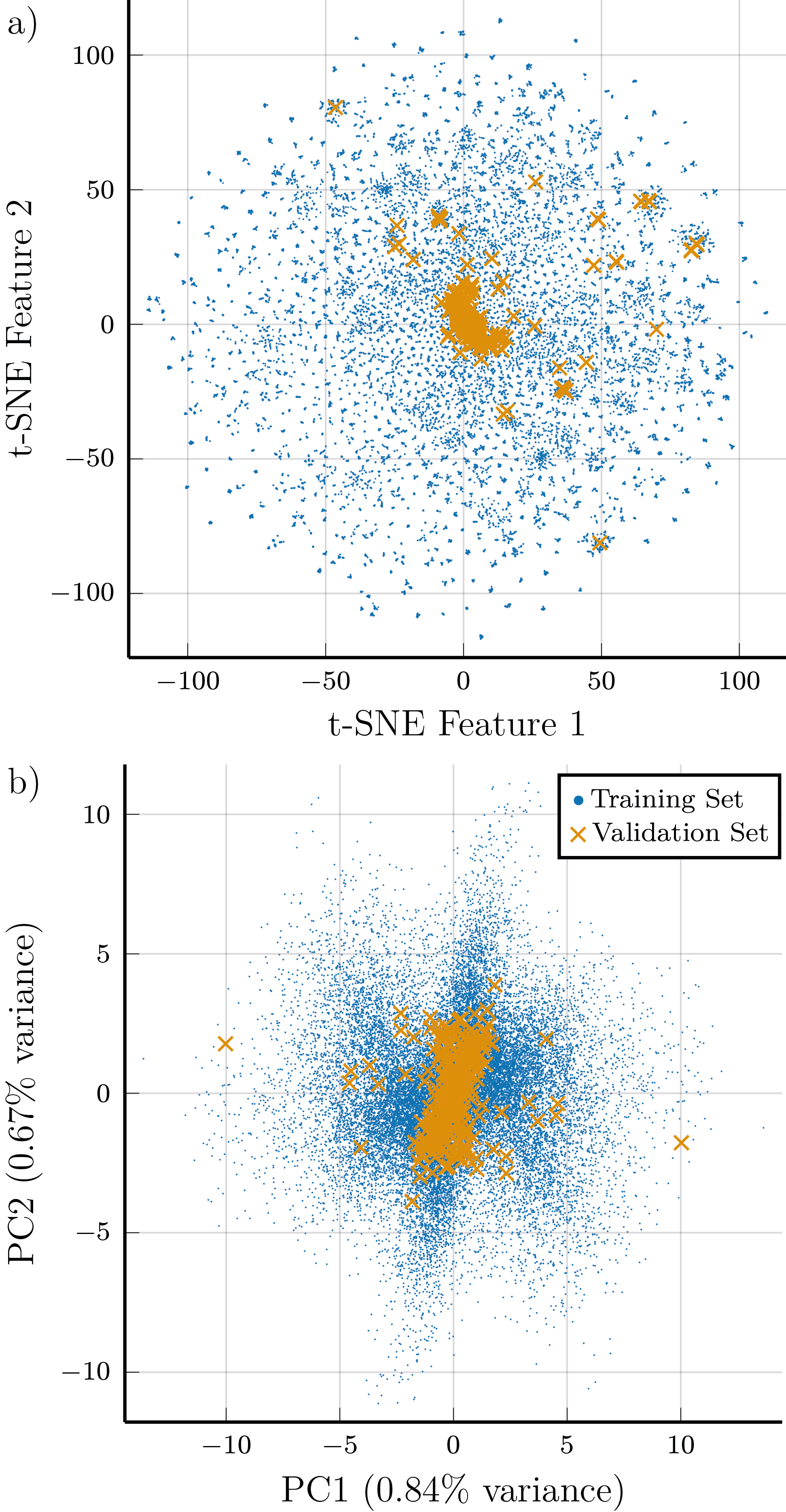}
    \vspace{-5pt}\caption{Low-dimensional representations of the chemical reaction space spanned by the radical-extended training set and the $C_4^{I, 0.01}$ validation set, as seen by the KPM reaction difference descriptor. a) t-SNE representation. b) PCA representation, using the two highest-variance PCs.}
    \label{fig:ml-kpm-analysis-tsne}
\end{figure}

Accounting for correlations between features better represents how the NNs within KPM see the reaction descriptors during training. We therefore use t-distributed stochastic neighbor embedding (t-SNE) to map the high-dimensional KPM descriptors into a two-dimensional space.\cite{vanderMaaten2008} This allows for visualisation of the representative chemical reaction space spanned by the training set, as well as enabling the space spanned by the validation set to be overlaid in order to evaluate the extent of overlap of these distributions. The descriptors for both training and validation sets were concatenated into a single dataset (because t-SNE learns a `single-shot' non-parametric mapping that cannot be reused for the validation set) and principal component analysis (PCA)\cite{Jolliffe2016} was used to reduce the dimensionality of this dataset to 100 (a noise reduction technique used in the original formulation of t-SNE).\cite{vanderMaaten2008} The t-SNE mapping employed a perplexity parameter of 50, and the two component datasets were split after being dimensionally reduced in order to produce the `map' of chemical space shown in Fig. \ref{fig:ml-kpm-analysis-tsne}a.

This mapping shows the reactions from the validation set entirely contained within the space encompassed by the training set; as such, the validation set reactions should be entirely in-distribution and their $E_a$ predictions should be accurate which, again, is not observed. To check that t-SNE was learning a valid low-dimensional representation of the chemical reaction space spanned by the datasets, the nearest-neighbours (in 2D t-SNE space) of each of the validation set reactions were taken from the training set reactions. These were used to establish whether similar reactions were being correctly grouped in t-SNE feature space. A selection of these nearest-neighbour reactions from the high-density central region of Fig. \ref{fig:ml-kpm-analysis-tsne}a are shown in Table \ref{tab:ml-kpm-analysis-tsne_neighbours}.

These nearest-neighbours initially appear sensible; in all four cases there are clear similarities between each selected reaction and its nearest neighbour. Each reaction shares a similar value of $\Delta H_r$ as its nearest neighbour from the training set, with the maximum difference being around 3 kcal/mol. However, the target activation energies for these reactions can differ greatly, by up to 70 kcal/mol. Closer inspection reveals that, while the reactant/product structures for selected reactions and their nearest-neighbours are typically similar, the underlying details of the reaction can be very different. For example, for reaction A, the selected reaction involves barrierless insertion of a carbene into a \ce{C-H} bond, while its nearest neighbour reaction inserts a carbene into a \ce{C-O} bond in a heterocycle. Similarly, for reaction C, the selected reaction involves dissociation of a hydrogen radical while its nearest-neighbour requires dissociation of molecular hydrogen from a stable hydrocarbon. 

\begin{table*}[t]
    \centering
    \includegraphics[width=0.9\textwidth]{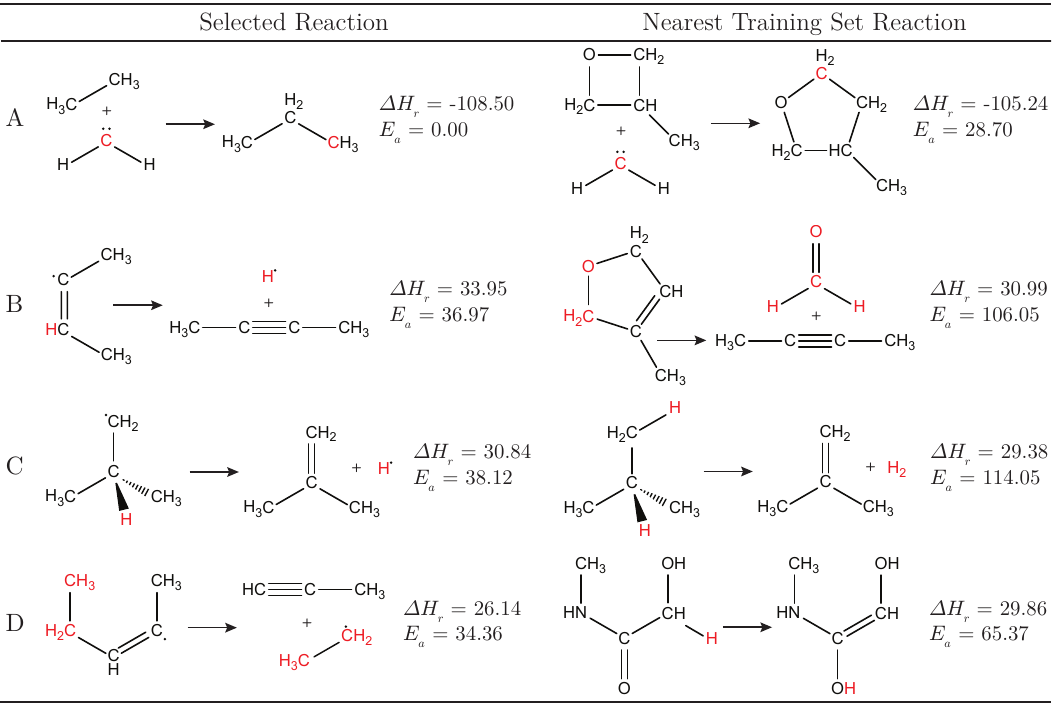}
    \caption{Selected reactions from the $C_4^{I, 0.01}$ validation set and their nearest neighbours in 2D t-SNE space from the radical extended training set. Atoms directly involved in each reaction are highlighted in red. All energy values are given in units of kcal/mol.}
    \label{tab:ml-kpm-analysis-tsne_neighbours}
\end{table*}

This nearest-neighbour analysis shows that, while t-SNE with the KPM descriptor captures \emph{structural} and \emph{enthalpic} similarity between reactions, it fails to account for the nature of the reaction. This either indicates that t-SNE is over-emphasizing particular descriptor features when constructing the 2-D representation of chemical reaction space, or that sufficient information is not present in the descriptors to enable adequate discrimination between reactions by t-SNE.

To better understand the mapping generated by t-SNE, we turned to an alternative dimensionality-reduction approach, namely PCA. We note that t-SNE focuses on preserving the structure of local clusters at the expense of preserving larger distances in the underlying high-dimensional space; by contrast, PCA is a transformation approach that better preserves distances between data points at both large and short distances (strictly, when data is transformed using the same number of principal components as input dimensions).\cite{Jolliffe2016} PCA can be used as a dimensionality reduction technique by selectively ignoring the principal components (PCs) that explain the least variance within the data. Here, a PCA model was fit using the 1025-dimensional KPM descriptors of the training set reactions, and the reactions of both training set and validation set were subsequently projected into the space spanned by the two PCs that captured the greatest variance; the results are shown in Fig. \ref{fig:ml-kpm-analysis-tsne}b.

This initially appears to confirm the conclusion from the t-SNE analysis --- that the validation set reactions lie within the distribution spanned by the training set. Specifically, we find the large majority of validation set reactions to lie near the center of the chemical reaction-space spanned by the training set; a line of symmetry is also present, which is expected due to the inversion of the forward reactions to create the respective backward reactions. However, the PCA analysis does reveal a stark failure of KPM's reactant-product difference descriptors --- the variances within the high-dimensional dataset that are captured by each PC (as shown in the axis labels of Fig. \ref{fig:ml-kpm-analysis-tsne}b) are extremely low, indicating that there is essentially no correlation between the individual features of the KPM descriptor. For completeness, this analysis was extended to the first ten PCs, which revealed that all together, these PCs only account for 4.81\% of the overall variance of the descriptor.

This is a crucial finding --- the inability of PCA to identify correlation between features of KPM's reactant-product difference descriptor explains much of the trained KPM models' inability to generalise the predictive accuracy of its test set results to the $C_4^{I, 0.01}$ validation set. The lack of correlation means that the NNs that make up KPM's predictive ensemble learn very little from the KPM descriptor itself, instead primarily relying on $\Delta H_r$ to drive their predictions. This reliance on $\Delta H_r$ is a major contributing factor to the high errors seen previously in Fig. \ref{fig:ml-kpm-ood-test_only}; learning directly from $\Delta H_r$ with only minor corrections contributed by the Morgan difference fingerprint means that reactions with similar values of $\Delta H_r$ appear similar to the model, as evidenced by the results in Table \ref{tab:ml-kpm-analysis-tsne_neighbours}. Including new radical reactions with similar values of $\Delta H_r$ to previously-seen reactions, but which often proceed through very different mechanisms and TSs, inevitably leads to inaccuracies in predicting activation energies which are intrinsically dependent on the structure of the TS. In other words, the descriptors used in KPM do not accurately represent the \emph{path} a reaction takes. 

KPM includes $\Delta H_r$ as part of its descriptor in accordance with prior literature, and this approach has been echoed in further contemporary work since.\cite{Ismail2022,Atwell2022,Lalith2024,Spiekermann2022,Vadaddi2024} It is a sensible choice, stemming from the Brønsted-Evans-Polanyi relation that correlates $\Delta H_r$ with $E_a$ in a variety of applications.\cite{Bronsted1928,Evans1938} The approach taken in KPM therefore mimics a $\Delta$-ML model, where $\Delta H_r$ is modified to approximate $E_a$ by means of learning from the remaining input features. Care must be taken though to ensure that these features are useful for learning; following on from our PCA analysis, we note that recent work has used SHapley Additive exPlanations (SHAP)\cite{Lundberg2017} to examine the relative importance of each feature on resulting predictions, finding that $\Delta H_r$ was considerably more useful than any of the features from a Morgan fingerprint-derived reaction descriptor.\cite{Cao2024} While the authors did not attempt to use their model on an out-of-distribution dataset such as the $C_4^{I, 0.01}$ validation set used here, we anticipate that they would run into the same roadblock of generalisability.

\subsection{Chemprop}

With better insight into the KPM model's lack of generalisability, we subsequently sought to benchmark alternative ML models in order to determine whether descriptors created by the difference between reactants and products --- but based on information other than Morgan fingerprints --- discard too much information to be useful in generalization. Chemprop is a D-MPNN originally formulated for molecular property prediction.\cite{Yang2019} It constructs one-hot encoded feature vectors for atoms and bonds and passes this information along molecular graphs to create a representation with an understanding of the chemical composition of a given chemical species. By adding so-called RDKit fingerprints --- a descriptor similar to Morgan fingerprints with a focus on identifying variable-size subgraphs rather than bonding patterns within a fixed radius\cite{Landrum2024} --- Chemprop has been shown to perform well at a wide range of regression and classification tasks.

Chemprop was extended to describe reactions for the purpose of $E_a$ prediction by Grambow \latin{et al.}, by means of an atom-mapped difference of the learned representation of a reaction's reactants and products.\cite{Grambow2020Chemprop} In their work, this model was trained on the same dataset as the original formulation of KPM, resulting in a test set RMSE of $3.4 \pm 0.3$ kcal/mol --- very similar to the RMSE that the authors of KPM originally achieved.\cite{Ismail2022}. While Grambow \latin{et al.} did not include $\Delta H_r$ as a feature explicitly, it was added as a secondary prediction target, allowing for implicitly learning its relationship with $E_a$.

To enable direct comparison between Chemprop and KPM in predicting activation energies of the $C_4^{I, 0.01}$ validation set, we retrained Chemprop from scratch using the same radical-extended training dataset as KPM. In accordance with the original training protocol for reaction-based Chemprop, this involved a two-stage process wherein the model was initially trained on the more abundant, lower accuracy version of this dataset comprising 36,778 reactions from DFT with the B97-D3 exchange-correlation functional, followed by training on the smaller but higher accuracy version of the dataset with 26,656 reactions at the $\omega$B9X-D3 level (the latter of which was used to train KPM). This has the effect of giving the model a greater appreciation of chemical reaction space from the first dataset, while elevating its predictions to the level of the second. Training was performed using the optimal hyperparameters from ref. \citenum{Grambow2020Chemprop}.

This resulted in an ensemble of ten unique committee members, all trained on different, although overlapping, areas of chemical reaction space defined by performing 10-fold cross validation on the full training dataset. The RMSEs of each of the ensemble members' predictions on their own test sets were averaged to give a final ensemble RMSE of $5.65 \pm 0.51$ kcal/mol. While this is greater than the RMSE of the original ensemble of Grambow \latin{et al.},\cite{Grambow2020Chemprop} this increase in error mirrors the increase seen when retraining KPM, where introduction of the additional dataset of radical reactions caused overall prediction quality to worsen. The new Chemprop ensemble was then used to predict activation energies and reaction enthalpies of the $C_4^{I, 0.01}$ validation set. As in the KPM ensembles employed previously, these properties were predicted for each reaction by every ensemble member and averaged to obtain the final prediction, with predictive uncertainties represented by the standard deviation between the predictions of the members. 

\begin{figure}[h!]
    \centering
    \includegraphics[width=240.71031pt]{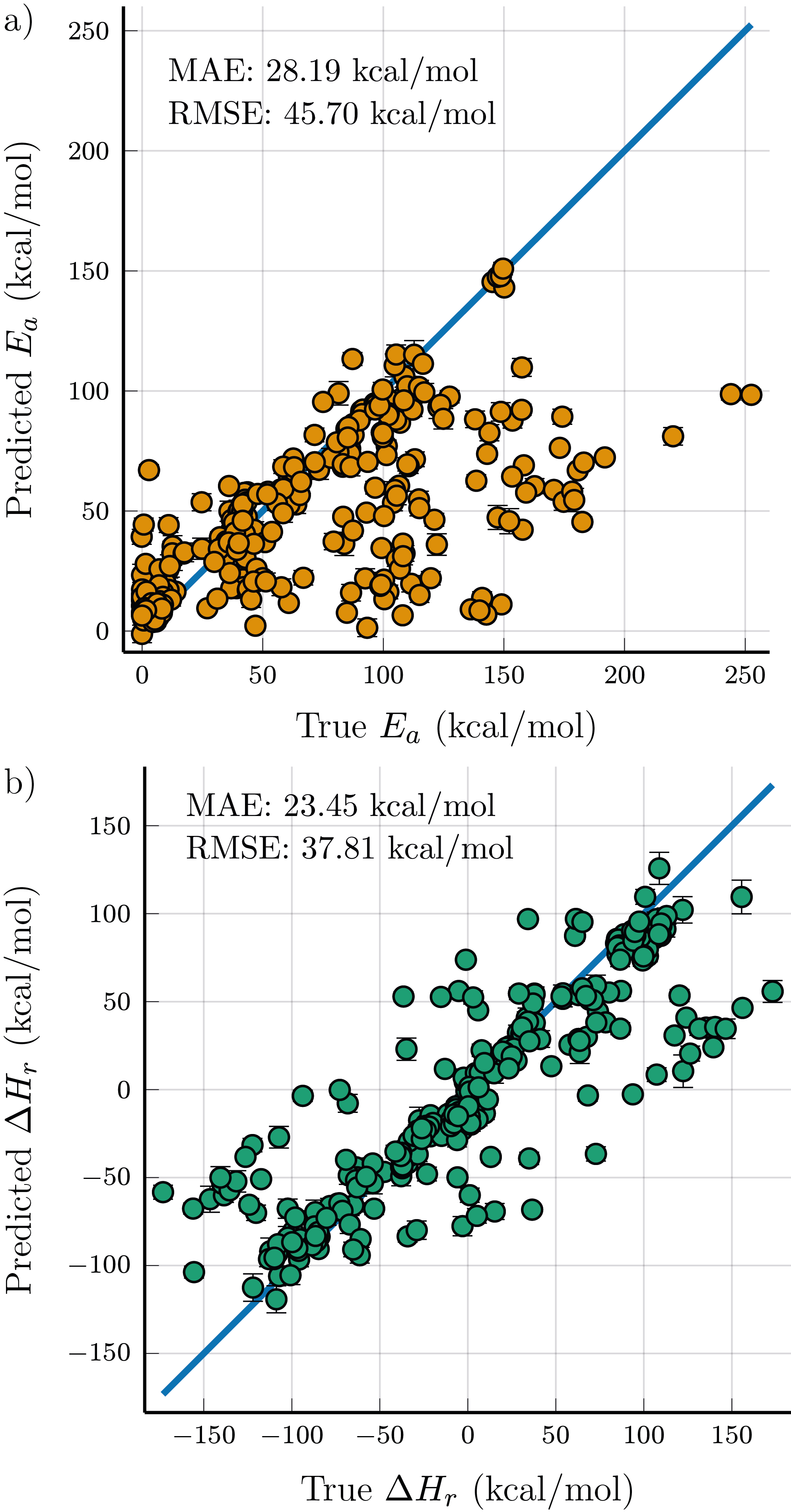}
    \caption{Correlation plots for a) $E_a$ and b) $\Delta H_r$ predictions by Chemprop for the reactions of the $C_4^{I, 0.01}$ validation dataset, when trained from scratch on the radical-extended dataset.}
    \label{fig:ml-bmk-chemprop}
\end{figure}

Correlation plots for these predictions are shown in Fig. \ref{fig:ml-bmk-chemprop}. We found the performance of this radical reaction-aware Chemprop ensemble to be broadly comparable to the results seen with KPM, as confirmed in the given error metrics. For KPM (Fig. \ref{fig:ml-kpm-ood-test_only}) there was little bias towards over- or under-estimation of activation energies, whereas Chemprop broadly underestimates $E_{a}$. This bias is not mirrored in Chemprop's $\Delta H_r$ predictions however; the reactions lying far from the identity line are equally distributed on either side (and in fact possess a rough line of symmetry that reflects the forward/backward reaction pairs). This is promising, suggesting that Chemprop is capable of transferring some of its predictive accuracy from training to the out-of-distribution validation reactions.

\begin{figure}[h]
    \centering
    \includegraphics[width=240.71031pt]{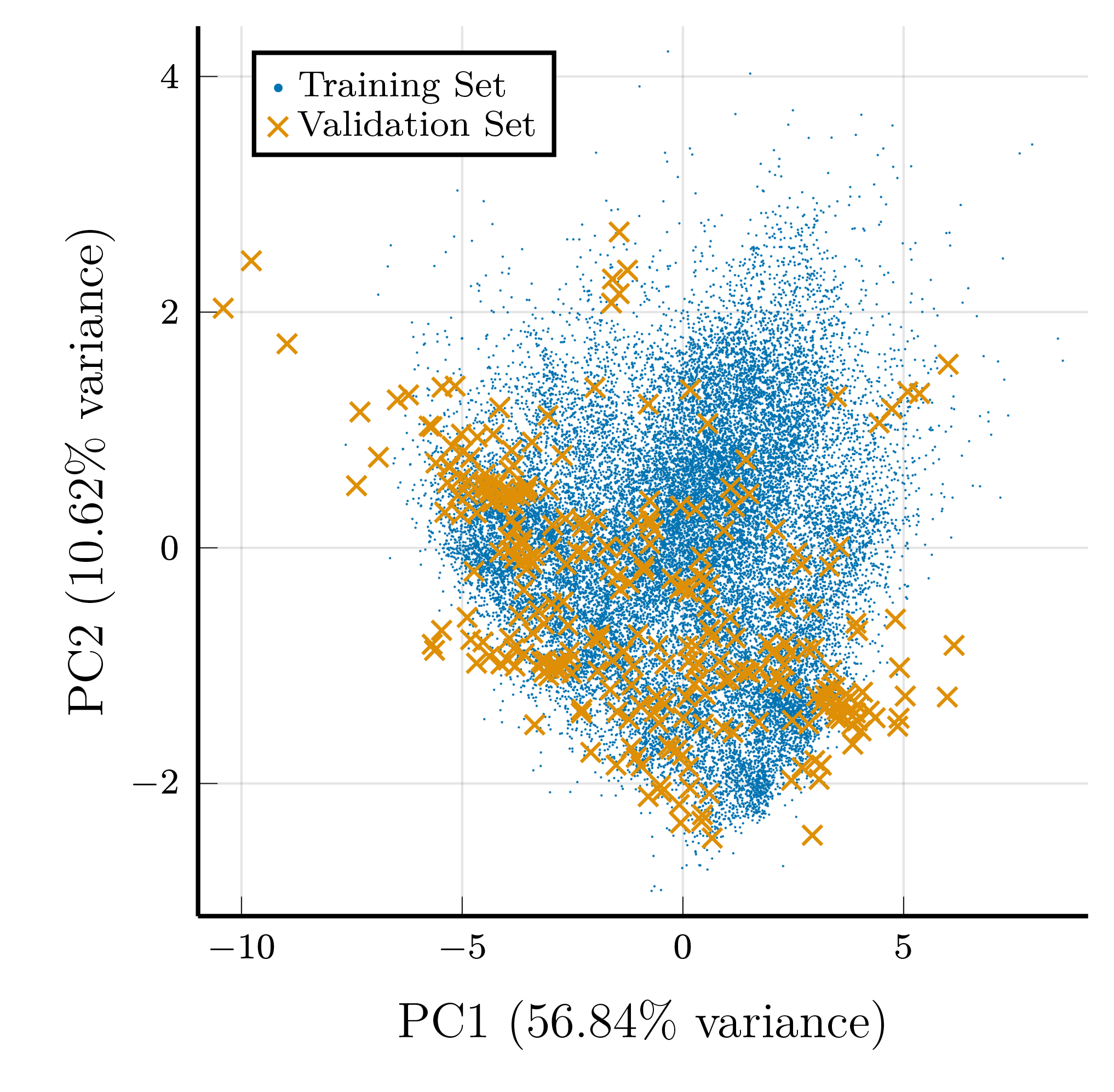}
    \caption{Low-dimensional PCA representation of the chemical reaction space spanned by the radical-extended training set and the $C_4^{I, 0.01}$ validation set, as seen by the reaction difference descriptor of the best performing model in the Chemprop ensemble.}
    \label{fig:ml-bmk-chemprop-pca}
\end{figure}

To develop better insight into Chemprop's approach, we extracted the final learned representation of its fingerprints from the penultimate NN layer and subsequently used this 1800-dimensional fingerprint in PCA analysis, following the same procedure as outlined above (because each NN ensemble member learns its own representation of a reaction, only the NN with the lowest individual RMSE was used in this test). The resulting 2D representation of chemical reaction-space --- projected onto the PCs encapsulating the largest variance --- is shown in Fig. \ref{fig:ml-bmk-chemprop-pca}. These results confirm that Chemprop is much more effective at encoding learnable information than KPM, with over half of the total variance in the learned descriptor being captured by the first PC. The relationship between the spaces spanned by the reactions of the validation set and the training set is correspondingly much clearer than in Fig. \ref{fig:ml-kpm-analysis-tsne}b, with Chemprop learning a representation that is able to generalise to some of the validation set. However, many of the validation set reactions lie on the fringes of the learned chemical reaction space, suggesting poorer predictive accuracy in these areas.

Despite the improvement in useful, learnable information in Chemprop compared to KPM, Chemprop is still unable to yield accurate predictions of the activation energies of the validation set. This may be due to how both Chemprop and KPM describe reactions; they use a difference fingerprint between reactants and products. This approach emphasizes \emph{differences} between the two end-points, singling out the atoms and bonds that are modified over the course of a reaction. However, this also annihilates much of the information which could provide context to discriminate between otherwise similar reactions, as properties such as the structure of both the reactants and the products, as well as the charges and radical electron counts of their atoms, are discarded.

\subsection{NeuralNEB}

So far, we have shown that two reactant-product-difference NN architectures --- KPM and Chemprop --- are unable to generalise to a challenging validation set of chemical reactions; as such, it is possible that such approaches simply lack the necessary information about the TS of each reaction. However, a machine-learned interatomic potential (MLIP) trained on sufficient examples of reactive intermediates and TSs may be able to account for this gap in knowledge. In this case, methods such as CI-NEB must be used to approximate the TS structure and activation energy.

We therefore consider NeuralNEB, a pre-trained version of the PaiNN equivariant message passing graph neural network.\cite{Schreiner2022,Schutt2021} The authors of NeuralNEB found that PaiNN performed best in reproducing the activation barriers of a set of test chemical reactions when trained on the Transition1x dataset.\cite{Schreiner2022T1x} Transition1x takes the reaction dataset by Grambow \latin{et al.},\cite{Grambow2020} originally used for training KPM, and extends it for use with MLIPs. The authors regenerated the MEPs of each reaction from the Grambow dataset with CI-NEB and sampled molecular geometries from across the available convergence space of each MEP. They obtained a dataset of 9.6 million molecular configurations, with energies and forces calculated at the same level of theory as used in our $C_4^{I, 0.01}$ validation set. NeuralNEB is therefore ideal for determining if MLIPs can perform better at $E_a$ prediction than reaction difference descriptor NNs when trained on similar data.

\begin{figure}[h!]
    \centering
    \includegraphics[width=240.71031pt]{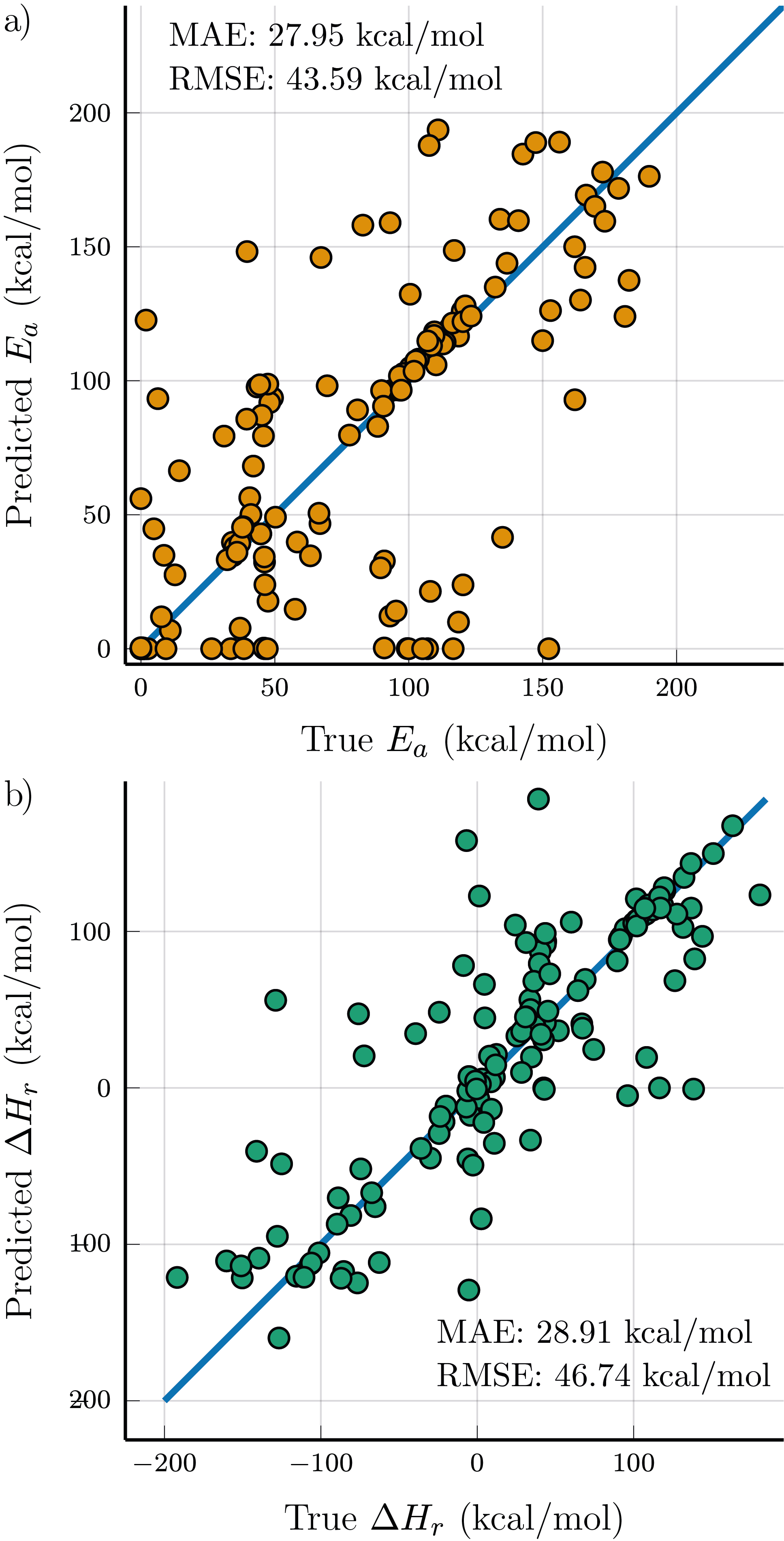}
    \caption{Correlation plots for a) $E_a$ and b) $\Delta H_r$ predictions by NeuralNEB for the reactions of the $C_4^{I, 0.01}$ validation set.}
    \label{fig:ml-bmk-corr_neuralneb}
\end{figure}

Since NEB calculations reveal entire reaction paths, not only a single activation energy, the multi-step reactions which were previously discarded from the validation set were reintroduced; even if it could only qualitatively reproduce these reaction paths, NeuralNEB could be used as part of a low-cost method for automatically separating these multi-step reactions into their constituent single-step reactions. To assess the performance of NeuralNEB, we therefore took the 381 reactions from network $C_4^{I, 0.01}$ that originally converged to a TS under DFT and subsequently subjected these to re-optimisation of reactant/product structures, followed by interpolation with IDPP and path-optimization through CI-NEB, using the Atomic Simulation Environment (ASE) software package as a driver for the calculations.\cite{Smidstrup2014,Larsen2017} MEPs were generated with the FIRE optimiser by running NEB until the maximum force experienced by any atom reached $0.1$ eV/$\angstrom$, then enabling the CI forces and continuing until this maximum force reached $0.04$ eV/$\angstrom$.\cite{Bitzek2006} 147 these reactions converged to a TS under NeuralNEB; their calculated activation energies and reaction enthalpies are shown in Fig. \ref{fig:ml-bmk-corr_neuralneb}. Selected NeuralNEB reaction MEPs are additionally compared against their DFT-level counterparts in Fig. S5.

We find that while NeuralNEB is capable of accurately predicting activation energies (and reaction enthalpies) for around half of these converged reactions, the remaining half are characterised poorly; the calculated MAE and RMSE are comparable to the errors predicted by Chemprop. However, it should be noted that the prediction task that MLIPs must undertake to predict MEPs and activation energies is more challenging than predicting a single value (as in KPM and Chemprop), instead demanding accurate prediction of large areas of a high-dimensional PES. This is emphasised by the number of reactions that failed to converge under CI-NEB, which failed due to numerical instability caused by large atomic forces, likely in areas of chemical space outside of NeuralNEB's training distribution. While the inclusion of the reactions of the secondary radical dataset of Grambow \latin{et al.}\cite{Grambow2020_raddata} within the Transition1x dataset might alleviate these blindspots, as with the previously investigated models, NeuralNEB is unable to use information on radical electron counts and charges within its description of the PES. As such, its predictive accuracy for radical reactions is still fundamentally limited, making it unsuitable for use within kinetic simulations involving radical species.

\section{Improving transferability}\label{sec::spahm}

The contemporary ML models tested above all failed to reliably predict the activation energies of the chemical reactions within the validation set of reactions from network $C_4^{I, 0.01}$, at least to an accuracy that would be usable within chemical kinetics simulations. A large part of this likely stems from these reactions containing many open-shell species. While Chemprop does encode information about atomic spin multiplicity in the feature vector that is passed between atoms in a molecule, and the resulting learned reaction descriptors are correlated and usable for the task at hand, it is likely that the subtractive approach used in creating these descriptors annihilates much of the useful contextual information that would enable a better understanding of radical species and reactions. Meanwhile, MLIPs like NeuralNEB rely on input geometries of species to evaluate the PES; having no information on spin or electron counts makes it unlikely that these models could learn how to properly characterise reactions involving radical species. Some newer models are beginning to emerge that may eliminate this deficiency, but none have been applied to the prediction of reaction energetics at this moment.\cite{Anstine2024, Bloore2021}

Based on the studies above, there are therefore three key areas of improvement for new ML models for $E_a$ prediction:

\begin{itemize}
\item \textbf{Maximizing information exploitation:} Models that rely on evaluating differences between products and reactants to obtain descriptors that solely represent bond changes may yield accurate predictions within a narrow slice of chemical reaction space, but sacrifice vital contextual information which could enable greater transferability. Models that predict reactive properties directly from reactants and products should therefore avoid featurising reactions using differences if transferability or generalisability is important.

\item \textbf{Accounting for spin:} While many models can accurately predict properties of reactions between closed-shell species, neglecting spin multiplicity can lead to erroneous predictions for reactions of open-shell species. Any model hoping to correctly predict the properties of such reactions should be `spin-aware', preferably building spin into the reaction descriptors. 

\item \textbf{Including TS Information:} While information about the true TS of a reaction is typically unknown to ML models based on reactant-product difference descriptors alone, providing a model with a TS \textit{approximation} could potentially enhance predictive accuracy.\cite{Spiekermann2022} Interpolation methods such as IDPP provide one way of obtaining approximate TS information,\cite{Smidstrup2014} which could subsequently be used to provide a better description of the path taken through chemical reaction-space.
\end{itemize}

Recently, a new representation of local atomic environments was developed which can better satisfy these criteria. This representation --- the global spectrum of approximated Hamiltonian matrices (SPA$^\textrm{H}$M) --- computes the occupied-orbital eigenvalues of an approximate `guess' Hamiltonian matrix for a molecular system.\cite{Fabrizio2022} These guess Hamiltonians are typically used as the starting point for self-consistent field iterations in DFT calculations; they are cheap to compute, but encode information about the system charge and spin, alongside a translationally- and rotationally-invariant geometry description. SPA$^\textrm{H}$M was shown to enable learning of molecular properties across a range of charge and spin states with accuracy greater than other contemporary global descriptors, such as SLATM (spectrum of London and Axilrod-Teller-Muto).\cite{Huang2020}

From SPA$^\textrm{H}$M, two further local representations have been developed --- SPA$^\textrm{H}$M(a) for encoding local atomic electron density, and SPA$^\textrm{H}$M(b) for encoding electron density of bonds surrounding each atom.\cite{Briling2024} These descriptors use the electron density calculated from guess Hamiltonians, decomposing it into atom-centered or bond-centered contributions. Both representations were shown to further improve predictive accuracy in ML models for problems related to spin and charge. SPA$^\textrm{H}$M(b) therefore stands to be an excellent choice of descriptor for predicting \textit{reactive} properties; it wraps knowledge about spin and charge into its representation, while locally encoding electron density of chemical bonds.

Here, in seeking to enhance accuracy and transferability of activation-energy prediction, we developed a new CNN architecture using SPA$^\textrm{H}$M(b) descriptors alongside approximate TS data obtained from geometry interpolation. As we show below, this approach reduces errors for the $C_4^{I, 0.01}$ validation set over all the methods tested above, albeit with some caveats discussed below.

\subsection{Reaction Interpolation}

We hypothesize that incorporating approximate TS geometries into a ML model could improve the accuracy and transferability of predictions of reactive properties; as such, we required a method for generating approximate TS geometries. Here, we sought to evaluate the performance of geometric interpolation schemes for TS approximation. However, we note our emphasis on using these interpolation strategies as part of an accelerated ML workflow, suggesting that any selected method for TS approximation must be computationally-efficient, broadly applicable, and employ purely geometric information. 

With these criteria, we sought to compare the performance of (i) linear Cartesian interpolation and (ii) IDPP,\cite{Smidstrup2014} as implemented in ASE, and (iii) geodesic interpolation\cite{Zhu2019} in approximating TS geometries. To evaluate performance, each of these three interpolation schemes was used to generate an approximate TS for each reaction in the radical-extended training dataset constructed previously. To determine the approximate TS geometry from each interpolation - while also maintaining the low computational cost of these interpolation methods - the highest energy geometry along the interpolated MEP (as characterised by GFN2-xTB calculations) was taken to be the approximate TS for the given reaction and interpolation scheme.\cite{Bannwarth2019}

The approximate TS geometries for each reaction, generated by each interpolation method, were then compared against their respective TS geometries in the DFT dataset. For this comparison, we chose to use the smooth overlap of atomic positions (SOAP) representation to construct local descriptions of the environments of each atom in both the DFT TS and the interpolated TS.\cite{Bartok2013,De2016} SOAPs are commonly used in ML models describing electronic structure, due to their ability to accurately encode regions of local atomic geometry into readily learnable vectors.\cite{Drautz2019} In this case though, each atom's local SOAP representation in the interpolated TS was compared against its counterpart in the DFT TS to calculate a distance metric in SOAP descriptor space, as follows:

\begin{equation}\label{eqn:ml-spahm-soap_dist}
    r_{\textrm{SOAP}} = \frac{1}{n_a}\sum_{i=1}^{n_a}\left \| \mathbf{S}_{i}^{\textrm{TS}} - \mathbf{S}_{i}^{\textrm{interp}} \right \|^2.
\end{equation}

\begin{figure*}[t]
    \centering
    \includegraphics{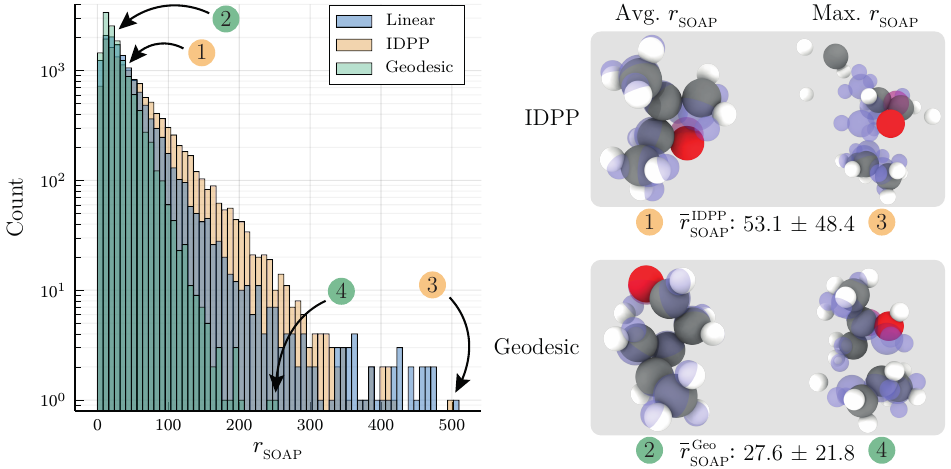}
    \caption{Distributions of $r_{\textrm{SOAP}}$ between exact TSs of the reactions in the radical-extended training set and their interpolated approximations. Included for the IDPP and geodesic interpolation methods are examples of approximate TS geometries at average and maximum SOAP distance, with their corresponding exact DFT TSs overlaid in blue. The statistical averages of the distributions for these methods are shown as $\overline{r}_{\textrm{SOAP}}^{\textrm{IDPP}}$ and $\overline{r}_{\textrm{SOAP}}^{\textrm{Geo}}$ respectively.}
    \label{fig:ml-spahm-interp-dists}
\end{figure*}

\noindent Here, $n_a$ is the number of atoms in the TSs, but $\mathbf{S}_{i}^{\textrm{TS}}$ and $\mathbf{S}_{i}^{\textrm{interp}}$ are each SOAP vectors of length $n_s$ from the $n_a \times n_s$ matrices $\mathbf{S}^{\textrm{TS}}$ and $\mathbf{S}^{\textrm{interp}}$. While this geometric similarity metric lacks the interpretability of metrics such as positional root mean squared deviation (RMSD), which operates in standard units of distance, it is translationally and rotationally invariant, allowing it to be much more robust to movements of groups of atoms. The distributions generated by this metric are shown in Fig. \ref{fig:ml-spahm-interp-dists}. 

This comparison of DFT TS structures and interpolated TS structures in the SOAP descriptor space reveals that geodesic interpolation significantly outperforms the other interpolation schemes, with the distribution of $r_{\textrm{SOAP}}$ strongly skewed towards lower values when compared to either linear or IDPP interpolation. Linear interpolation performs the worst overall, with comparatively many approximate TSs at high values of $r_{\textrm{SOAP}}$ due to its tendency to let atoms intersect during interpolation. IDPP creates less of these high-distance TSs, but those it does create tend to look like structure 3 in Fig. \ref{fig:ml-spahm-interp-dists}. This explosion of atoms is typical of IDPP edge cases where geometry optimisation under the IDPP potential fails numerically. By contrast, even the worst-performing geodesic TS maintains a sensible structure that, while different from its DFT counterpart, still broadly matches the connectivity and placement of its atoms. Tabulated results and further examples of low-, average- and high-$r_{\textrm{SOAP}}$ geometry differences are given in \textit{Supporting Information} Tables S1 and S2, respectively. 

Overall, our results show that geodesic interpolation allows for the generation of sufficiently accurate MEPs that, when combined with energy evaluation at the GFN2-xTB level, yield approximate TS geometries with only minor conformational differences compared to their respective DFT-level geometries across the vast majority of reactions studied. It therefore represents an excellent choice of method for generating approximate TS geometries for ML models that predict reactive properties, as we now discuss.

\subsection{Convolutional Neural Network Architecture}

Armed with geometries for the reactants, approximate TS, and products of reactions, and with the SPA$^\textrm{H}$M(b) descriptor of chemical bonds around local atomic environments, we can begin to learn reactive properties with a suitable ML model. Rather than subtracting these descriptors for a given reaction in the style of KPM and Chemprop, here we choose to learn the optimal combination of each feature across the three geometries with a convolutional neural network (CNN). CNNs are common in many modern image recognition techniques, where large 2D pixel grids are distilled down to smaller representations which maximize the effect of important parts of the image using convolution filters (sometimes called convolution kernels).\cite{Gu2018, Aghdam2017} In a CNN, this convolution filter is a matrix of learnable weights, enabling models to learn the best way of performing this size reduction. This is typically followed by a pass through a nonlinear activation function in the same way that a hidden layer in a MLP might be, forming a single convolutional layer. 

To apply these techniques to SPA$^\textrm{H}$M(b) descriptors, the local representations of each atom in the reactants, TS and products were fed through such a convolutional layer, enabling prediction of per-atom properties. However, such properties are not available for training; $E_a$ and $\Delta H_r$ are intrinsically properties of a reaction as a whole. Instead, the CNN must learn its own local contributions to these target properties, so that they can be summed to create the global properties of interest.

\begin{figure*}[t]
    \centering
    \includegraphics{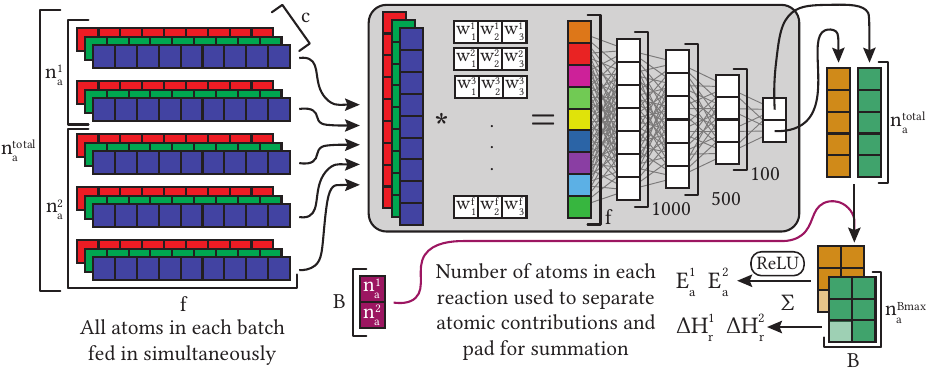}
    \caption{Representation of the architecture of the DCINet CNN. Vectors of $f$ features representing atoms in the reactants, TS and products (denoted here as red, green and blue channels $c$) undergo a depthwise convolution to obtain a single reaction descriptor of $f$ features. This is passed through a series of feed-forward layers to output atomic contributions to $E_a$ and $\Delta H_r$. The number of atoms in each reaction $n_a^i$ in a mini-batch of size $B$ is also provided as an input, allowing atomic contributions to target properties to be correctly split and zero-padded for summation.}
    \label{fig:ml-spahm-atoms-indivnet_eg}
\end{figure*}

The resulting CNN architecture, referred to henceforth as DCINet (depthwise convolutional individual-atom network), is shown in Fig. \ref{fig:ml-spahm-atoms-indivnet_eg}. Each atomic reactant, TS and product descriptor was considered to be a channel $c$ in a CNN. After normalising the features in each channel we chose to perform a depthwise convolution across channels, with each feature in the descriptor being convolved by an independent $1 \times 3$ learnable convolution filter. The output of this convolutional layer was therefore a single vector of length $f = 2,216$ for the SPA$^\textrm{H}$M(b) descriptors illustrated here. This was then fed into three feed-forward layers using the rectified linear unit (ReLU) activation function, reducing the descriptor in size until only two features remained in the output layer: the $E_a$ and $\Delta H_r$ output predictions.

However, in seeking to learn intrinsic reaction properties from individual atomic environment descriptors, different reactant/TS/product systems will have different numbers of atoms $n_{a}$. In order to correctly divide reactions into batches of size $B$ for use in the mini-batch gradient descent method commonly employed for training such models, the SPA$^\textrm{H}$M(b) descriptors for all atoms in the reactions of each mini-batch were stacked into a single $\left( n_a^\textrm{total} \times f \right)$ matrix (one per reactive channel), where $n_a^\textrm{total} = \sum_{i=1}^{B} n_a^i$. The number of atoms in each reaction within the mini-batch ($n_a^i$) was also fed into the CNN as an input.

Once a contribution to $E_a$ and $\Delta H_r$ has been predicted for each atom in the batch, the model then uses these input atom counts to redistribute these contributions among the input reactions, creating a $\left( n_a^{\textrm{Bmax}} \times B \right)$ matrix for each property (where $ n_a^{\textrm{Bmax}} = \underset{i \in B}{\textrm{max}} \; n_a^i$), padded with zeroes in the case that a given reaction has $n_a^i < n_a^{\textrm{Bmax}}$. Columns of these matrices are then summed to yield the final $E_a$ and $\Delta H_r$ predictions for each reaction in the batch, with the addition of a ReLU activation function after the $E_a$ summation to constrain its final values to be exclusively positive.

\subsection{DCINet Results}

SPA$^\textrm{H}$M(b) descriptors were generated for every reactant, TS and product in both the radical-extended training set and the $C_4^{I, 0.01}$ validation set using the Q-stack Python package.\cite{QStack} TS descriptors were calculated from both the true DFT NEB TS geometries and the geodesic interpolated TS geometries to compare the quality of model predictions against a best case scenario, where the interpolation exactly matches the true TS. 

We implemented DCINet in PyTorch and trained on a randomly shuffled 80\% split of the radical-extended reaction dataset, with 20\% left as a held-out test set.\cite{Ansel2024} Its parameters were optimised with the Adam stochastic optimiser using an initial learning rate of $1\times10^{-3}$, reduced by a factor of $0.3$ whenever the loss of the test set predictions failed to decrease for 8 epochs, down to a minimum of $1\times10^{-6}$.\cite{Kingma2017} The loss function was defined as a linear combination of the mean squared error (MSE) of $E_a$ and $\Delta H_r$ predictions:

\begin{multline}
    l\left( \mathbf{x}^{E_{a}}, \mathbf{y}^{E_{a}},\mathbf{x}^{\Delta H_r},\mathbf{y}^{\Delta H_r} \right) = \dfrac{\alpha}{B}\sum_{i=1}^{B} L \left( \mathbf{x}^{E_{a}}_i,\mathbf{y}^{E_{a}}_i \right) + \dfrac{1-\alpha}{B}\sum_{i=1}^{B} L \left( \mathbf{x}^{\Delta H_r}_i,\mathbf{y}^{\Delta H_r}_i \right)\\
    \textrm{where } L\left( \mathbf{x},\mathbf{y} \right) = \left\{ l_1,\ldots,l_B \right\}^T \textrm{, } l_n = \left( x_n - y_n \right)^2
\end{multline}

\noindent where $\mathbf{x}^{E_{a}}$ and $\mathbf{x}^{\Delta H_r}$ are vectors of activation energy and reaction enthalpy predictions, $\mathbf{y}^{E_{a}}$ and $\mathbf{y}^{\Delta H_r}$ are their respective target values, and $\alpha$ is a tuning parameter that allows for control over the contribution of the two targets towards the overall loss function. In this case, $\alpha$ was set to $0.9$, meaning that 90\% of the loss was controlled by the accuracy of the $E_a$ predictions. In general, a value over $\alpha=0.5$ should be used here as $E_a$ is the primary training target. The batch size $B$ was set to 200.

\begin{figure*}[t]
    \centering
    \includegraphics[width=0.9\textwidth]{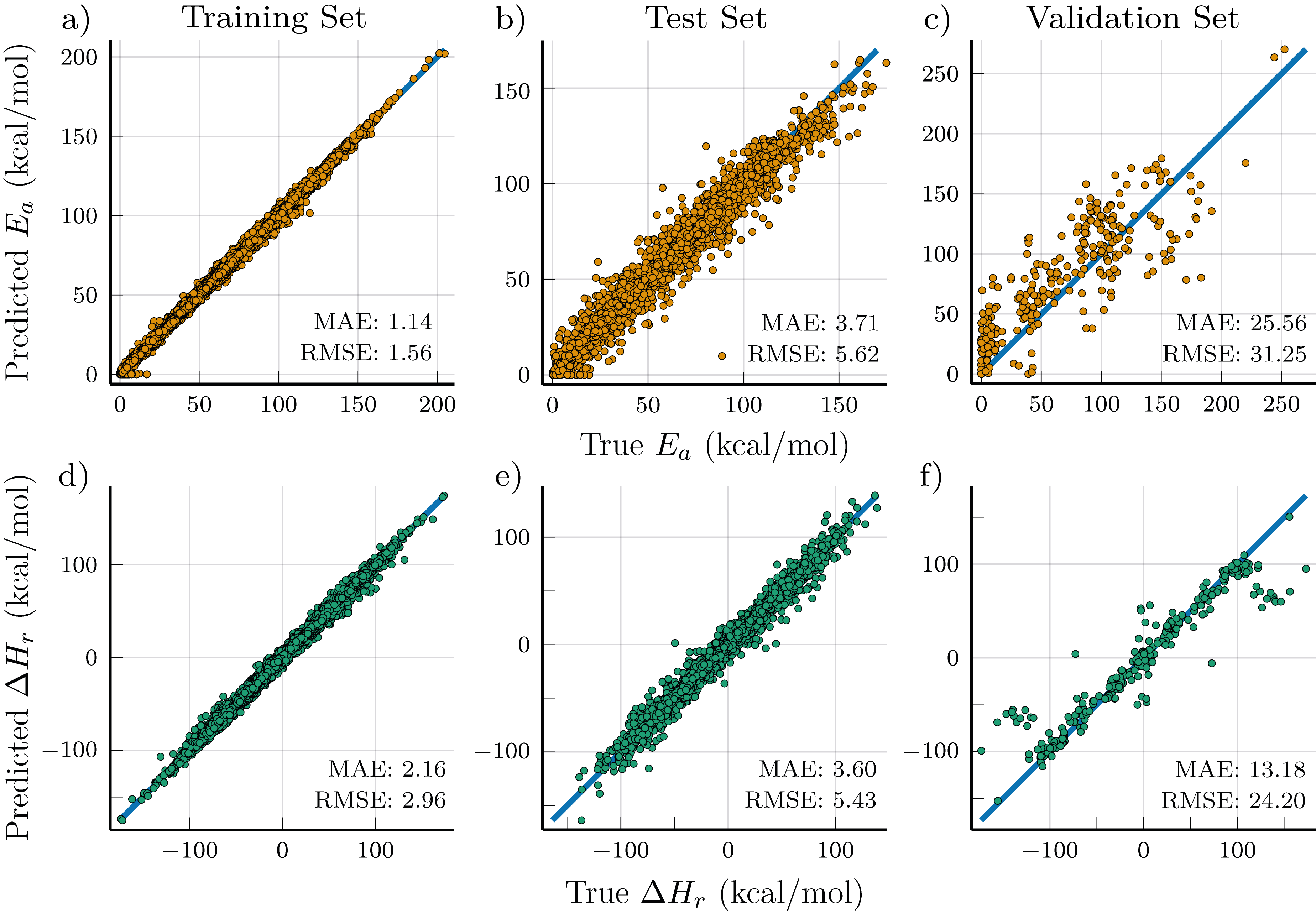}
    \caption{Correlation plots for a)--c) $E_a$ and d)--f) $\Delta H_r$ predictions of the DCINet model when applied to the training set, test set and $C_4^{I, 0.01}$ validation set respectively. All error metrics are given in units of kcal/mol.}
    \label{fig:ml-spahm-conv-corr}
\end{figure*}

The CNN was trained on an 80:20 training:testing set split of the radical-extended database, using the DFT-level TS geometries in both cases. Training proceeded until the loss of the test set predictions had stopped changing by more than $1\times10^{-4}$ for over 12 epochs, in total taking 298 epochs to complete. The correlation plots for the predictions of the training and test set reactions, along with the predictions of the model when applied to the reactions of the $C_4^{I, 0.01}$ validation set using geodesic-interpolated TS geometries, are shown in Fig. \ref{fig:ml-spahm-conv-corr}. Loss function plots visualising the progression of the learning rate are shown in Fig. S6.

The training and test set performance of this DCINet model matches that of both KPM and Chemprop, although it does not require averaging over a committee of models to achieve this accuracy. Relatively few reactions are poorly represented --- 81\% of the $\sim5300$ test set reactions fall under Ismail \latin{et al.}'s 6 kcal/mol error target. More impressive are the validation set results, which outperform any of the other models tested here in both $E_a$ and $\Delta H_r$ predictions. While these predictions are still not of sufficient accuracy to use in kinetic simulations (where errors are made exponentially worse during rate constant calculation), the incorporation of spin-aware descriptors and low-cost, approximate TS information provides a marked improvement in this challenging area of chemical reaction space. This is likely due to the additional context provided to DCINet extending the area of applicability of the training data. 

We can verify that the inclusion of approximate TS information has improved $E_a$ prediction by constructing a variant of the DCINet architecture that excludes TS information entirely. This means that the SPA$^\textrm{H}$M(b) descriptors for only two reactive channels --- the reactants and the products --- are convolved over to yield a reaction descriptor that enters DCINet's feed-forward layers. The resulting correlation plots for the same training, testing and validation sets are shown in Fig. S7. While $\Delta H$ predictions are only slightly worsened by this change, $E_a$ predictions are significantly worsened in all three datasets. This model performs similarly to Chemprop when predicting activation energies of the validation set, which indicates that convolution of only the reaction endpoint descriptors yields a similar lack of reactive context as the subtractive approach taken by Chemprop and KPM.

\begin{figure}[h]
    \centering
    \includegraphics[width=240.71031pt]{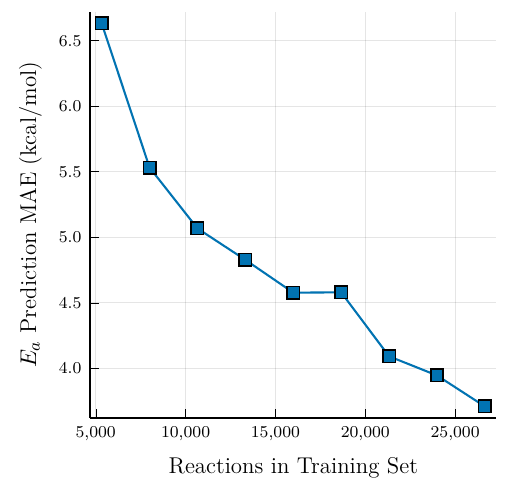}
    \caption{Learning curve for the DCINet model, showing the decrease in test set $E_a$ prediction MAEs as training set size is increased.}
    \label{fig:ml-spahm-learning-curve}
\end{figure}

The results in Fig. \ref{fig:ml-spahm-conv-corr} do not represent this model's true limit of applicability. We consider two avenues of improvement: the training set could be expanded, or the accuracy of the TS approximations could be improved. The potential effect of the former improvement can be determined by examining the learning curve shown in Fig. \ref{fig:ml-spahm-learning-curve}. The lack of a plateau in $E_a$ prediction errors as the training set size is increased indicates that the DCINet architecture continues to be capable of learning more if more training reactions were to be provided.

\begin{figure*}[t]
    \centering
    \includegraphics[width=0.95\textwidth]{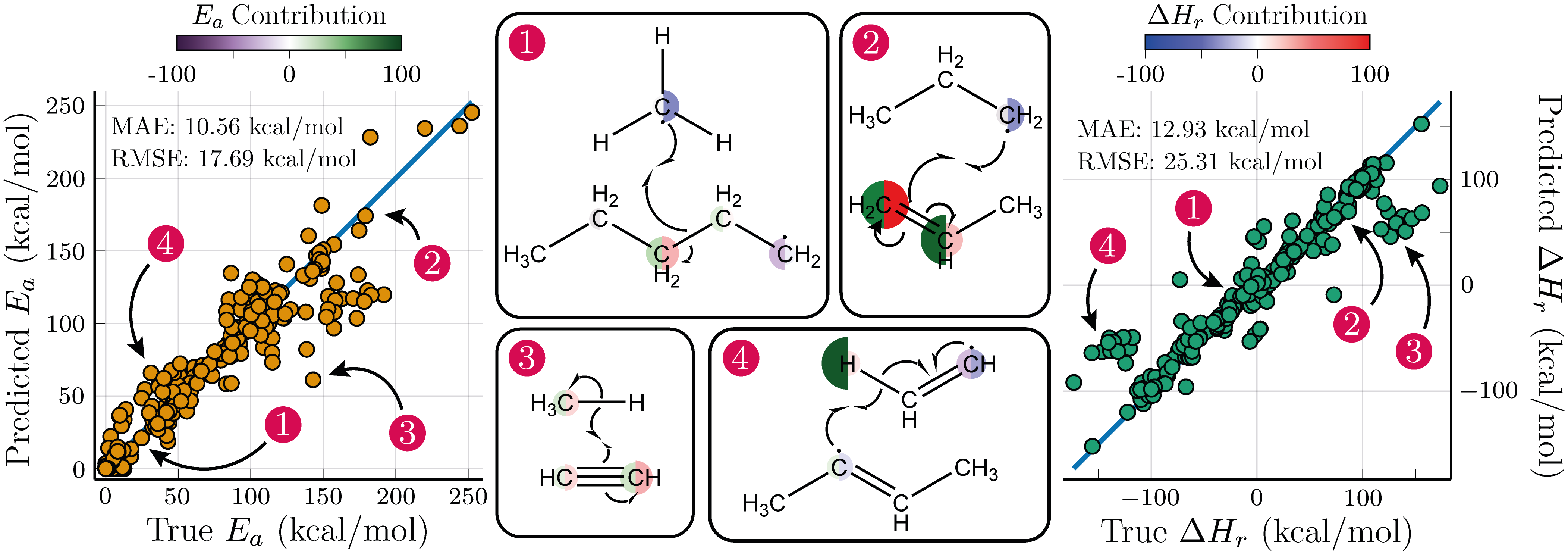}
    \caption{Correlation plots for $E_a$ and $\Delta H_r$ predictions of the DCINet model when applied to the $C_4^{I, 0.01}$ validation set when using true DFT-level TS geometries, representing a best-case scenario. Included are four examples of per-atom predictions of these quantities made by the model --- atoms are labelled with left and right semicircles according to the colour bars for $E_a$ and $\Delta H_r$ contributions respectively.}
    \label{fig:ml-spahm-atomwise}
\end{figure*}

As a proof-of-principle, we can also address the quality of the TS approximation by using the true DFT-level TS geometries when predicting reactive properties, instead of using approximate TS geometries. These predictions, shown in Fig. \ref{fig:ml-spahm-atomwise}, provide a significant improvement over the approximate TS predictions; the errors within the validation set decrease significantly. While this is not a feasible technique since obtaining these TS geometries requires performing the computationally expensive calculations that the model seeks to avoid, it is nevertheless informative, as these results represent the best-case scenario of predictions with this model architecture. Any remaining errors greater than the test set accuracy are due to reactions being truly out-of-distribution, where additional training data is required to obtain greater accuracy. 

While it would be useful to compare the positions of reactions in a low-dimensional latent space as in Figs. \ref{fig:ml-kpm-analysis-tsne} and \ref{fig:ml-bmk-chemprop-pca}, this is unfortunately not possible due to the atomwise predictions generated by the model, which prevent the formation of descriptors for entire reactions that can be dimensionally reduced. However, this architecture does create opportunities for other unique analyses; we can instead analyse the predicted \emph{atomic} contributions to $E_a$ and $\Delta H$. While such values have little physical meaning, they can nevertheless be interpreted and used for prediction of potential reactive sites and mechanisms. Fig. \ref{fig:ml-spahm-atomwise} also includes a handful of examples from the validation set reactions to demonstrate this capability.

The first two examples are well-predicted by the model, resulting in $E_a$ and $\Delta H_r$ prediction errors under 5 kcal/mol. Example 1 details an attack on a \ce{C-C} bond in a pentyl radical by a methyl radical, a process with relatively little $E_a$ required and which is neither endo- or exothermic, as while the reactants are two unstable radicals, so are the products. The model predicts a positive $E_a$ contribution from the two carbons between which a bond is broken, mainly centred on the carbon receiving a radical electron. The radical pentyl carbon contributes negatively to the $E_a$, possibly due to its perceived destabilising effect on the rest of the molecule being attacked. The radical methyl carbon does not contribute to the $E_a$ as it is already reactive and does not require additional energy to attack the \ce{C-C} bond, but it does contribute negatively towards $\Delta H_r$ as it becomes stably bonded to another carbon in the products. This is cancelled out in the final $\Delta H_r$ summation by an equivalent positive contribution on the pentyl carbon receiving the radical electron, which is destabilised by the same amount.

Example 2 depicts a very high-barrier process where a \ce{C=C} double bond is unevenly broken by an attack from a propyl radical, resulting in a 2-pentyl radical and a molecule of carbene. This requires the $\pi$-electrons from the bond to migrate to the terminal carbon of the propene, while a $\sigma$-electron is used in forming the new bond. This demands high $E_a$ contributions from the carbons being split, particularly on the one receiving the radical electron. Both carbons are also destabilised in the products, resulting in positive $\Delta H_r$ contributions, while the radical propyl carbon is stabilised by bond formation.

These examples show how the atomwise predictions of this model can be rationalised into reactivity and stability descriptors, similar to Fukui indices, a by-product of electron density calculation in DFT which allow chemical insight into how prone to electrophilic and nucleophilic attack individual atoms may be.\cite{Geerlings2003,Balawender1998} As such, Fukui indices are commonly used for predicting the reactivity of atoms in molecules.\cite{Olah2002} Likewise, these contributions are conceptual indicators which are based on (in this case, through SPA$^\textrm{H}$M(b)) electronic structure calculations, and while they have little physical meaning, can aid in the visualisation and prediction of complex reactive processes.

Likewise, this form of analysis can also be used to determine how and why this DCINet model fails to predict properties for other reactions; the latter two examples are reactions for which the model experiences high predicted errors for $E_a$ and $\Delta H_r$. Example 3 shows an attack by ethyne on a \ce{C-H} bond in methane. This is a highly unlikely reaction to occur in all but the most extreme conditions, but the DCINet model underpredicts both its $E_a$ 
and its $\Delta H_r$ by nearly 100 kcal/mol. Based on the accurately predicted examples, we would expect a larger positive $E_a$ contribution on the ethyne carbons, across which a $\pi$ bond is being broken, and greater destabilisation on the methyl carbon on the form of a larger positive $\Delta H_r$ contribution.

In example 4, both $E_a$ and $\Delta H_r$ are instead overpredicted. This reaction depicts a radical attack of a \ce{C-H} bond in a vinyl radical by a 2-butenyl radical. The model likely overpredicts the $E_a$ contribution of the departing hydrogen, and fails to account for the additional stabilisation gained by both the attacking and the attacked carbons, the former of which loses its radical electron while the latter participates in a strong triple bond.

These reactions may be poorly predicted for a number of reasons, perhaps the most obvious of which is simply that they lie too far outside of the chemical reaction space covered by the training set. Both contain hydrogen abstractions that either form or break \ce{C#C} triple bonds; these are poorly represented as only 75 radical reactions containing \ce{C#C} bonds exist in the training set, and heterolytic cleavage across \ce{C-H} bonds is similarly sparse. The validation set reactions that are predicted very well instead tend to involve \ce{C-C} single and double bonds being modified, making it likely that better prediction would require more examples of these rare reaction types.

\section{Conclusions}

To examine the potential hurdles that need to be overcome for ML models to accurately drive kinetic simulations of large CRNs, we have undertaken a comprehensive analysis of the performance of several contemporary ML models for chemical reaction property prediction. We defined a challenging validation dataset of reactions taken from an algorithmically-generated CRN for the pyrolysis of ethane. These reactions frequently contain free radical species, which are typically not covered by contemporary reaction datasets. We tested the applicability of three previously proposed ML models: KPM, Chemprop and NeuralNEB.\cite{Ismail2022,Grambow2020Chemprop,Schreiner2022} Although the training- and test-set performance of these models seems promising, none were transferable enough to yield activation energy predictions suitable for kinetic simulations. For example, KPM and Chemprop --- two models based on a reactant/product-difference descriptor --- likely do not know enough about the nature of the TS to distinguish between ionic and radical reaction pathways, leading to significant errors.

Based on these findings, we proposed a model that takes advantage of computationally inexpensive, approximate TS information to enhance transferability. We developed the DCINet architecture --- a CNN using the electronic structure-derived SPA$^\textrm{H}$M(b) descriptors to represent atomic environments across reactants, geodesic-interpolated TSs and products --- and showed that it achieves higher accuracy when applied to the validation set than any other model tested here. While still far from usable for chemical kinetics, we proved that even cheap interpolated TS information can dramatically improve models and that the prediction improves as the TS becomes more accurate.

We also discovered that the atomic contributions to reactive properties which DCINet produces as an intermediate step can be used as interpretable indicators for atomic chemical reactivity and stability when applied to well-predicted reactions. Conversely, when applied to poorly predicted reactions, the contributions which fail to be rationalised by chemical intuition indicate functional groups and reactive trends that are not present in the training data, highlighting which types of reactions require additional training data to improve the overall predictive accuracy. The DCINet architecture shows room for further improvement as the training set increases in size, indicating that such future additions would likely yield further increases in predictive accuracy.

Aside from the enrichment of the training data, there are further avenues by which DCINet could be improved in the future; the most obvious of these is by improving the accuracy of the TS estimation. While we showed that geodesic interpolation performs well across a diverse range of reactions, machine-learned solutions such as TSDiff have shown great promise when estimating TS geometries, and could therefore stand to increase the predictive accuracy of DCINet too.\cite{Kim2024} Additionally, the convolutional nature of DCINet enables the inclusion of snapshots along the reaction path other than the reactants, TS and products. Such snapshots could be generated through interpolation and added as extra channels in the convolution to provide further context to models about a reaction's movement across the PES. These potential improvements, coupled with the increased transferability already achieved here, may pave the way for substantially enhanced ML predictors of reactive properties.

\section*{Author contributions}

\textbf{Conceptualization}: Joe Gilkes and Scott Habershon. \textbf{Software, data curation and visualisation}: Joe Gilkes.
\textbf{Writing and formal analysis}: Joe Gilkes, Reinhard Maurer and Scott Habershon.
\textbf{Supervision}: Mark Storr, Reinhard Maurer and Scott Habershon.

\section*{Code and data availability}

The reaction-discovery code used to generate chemical reactions for further ML testing, 
\textit{Kinetica.jl}, is available on GitHub at \url{https://github.com/Kinetica-jl/Kinetica.jl}. \textit{KineticaKPM.jl} is similarly available on GitHub at \url{https://github.com/Kinetica-jl/KineticaKPM.jl}. Both packages are installable through the Julia language \verb|Pkg| package manager. Data will be made available through the University of Warwick research archive portal (wrap.warwick.ac.uk) upon acceptance.

\begin{acknowledgement}

The authors acknowledge funding from the Atomic Weapons Establishment, the EPSRC Centre for Doctoral Training in Modelling of Heterogeneous Systems (EP/S022848/1) at the University of Warwick, a UKRI Future Leaders Fellowship (MR/S016023/1 and MR/X023109/1), and a UKRI frontier research grant (ERC StG, EP/X014088/1). Computing facilities were provided by the Scientific Computing Research Technology Platform (SCRTP) of the University of Warwick.

\end{acknowledgement}

\begin{suppinfo}

The \textit{Supporting Information} contains further details on ML model training and performance, as well as details on the analysis of chemical reaction space spanned by datasets described above.

\end{suppinfo}

\newpage
\bibliographystyle{achemso}
\bibliography{bibliography}

\newpage
\appendix
\renewcommand{\thesection}{S\arabic{section}}
\renewcommand{\theequation}{S\arabic{equation}}
\renewcommand{\thetable}{S\arabic{table}}
\renewcommand{\thefigure}{S\arabic{figure}}
\renewcommand{\thepage}{\roman{page}}
\setcounter{figure}{0}
\setcounter{table}{0}

\begin{titlepage}
    \begin{center}
        \begin{LARGE}
            \textbf{Supporting Information:\\}
        \end{LARGE}
        \vspace{5mm}
        \begin{large}
            \textbf{Understanding and improving transferability in machine-learned activation energy predictors  \\}
        \end{large}
        \vspace{5mm}
        {\large Joe Gilkes$^{1,2}$, Mark Storr$^3$, Reinhard J. Maurer$^{1,4}$, Scott Habershon$^1$}\\
        \vspace{5mm}
        $^1$Department of Chemistry, University of Warwick, Gibbet Hill Road, CV4 7AL Coventry, UK
        
        $^2$EPSRC HetSys Centre for Doctoral Training, University of Warwick, Gibbet Hill Rd, CV4 7AL, Coventry, UK
        
        $^3$AWE Plc, Aldermaston, UK
    
        $^4$Department of Physics, University of Warwick, Gibbet Hill Road, CV4 7AL Coventry, UK

        \vspace{10mm}
        UK Ministry of Defence \copyright \, Crown Owned Copyright 2025/AWE
        \vfill
    \end{center}
\end{titlepage}

\section{Dataset Generation}

\begin{figure}[h]
    \centering
    \includegraphics{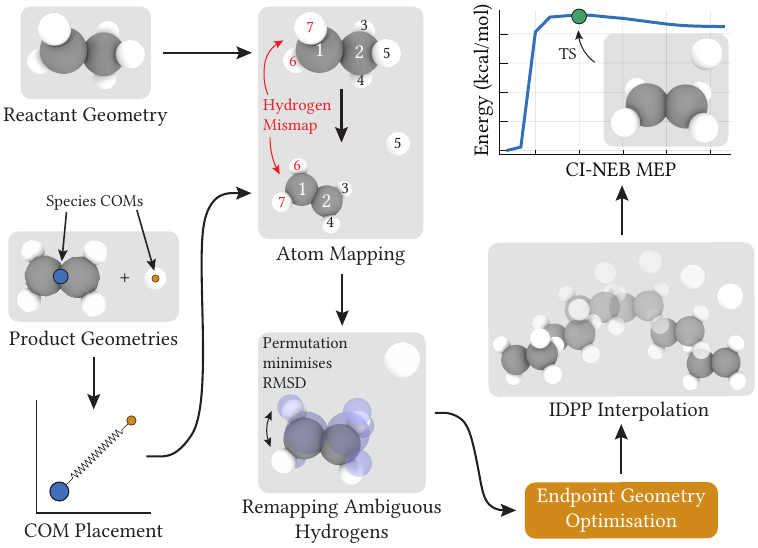}
    \caption{Overview of the workflow used to calculate converged MEP from isolated species geometries, performed on a reaction where a hydrogen atom dissociates from an ethyl radical to form ethene. If there are multiple reactants or products, they are placed by movement of their COM. Atomic indices are mapped to be consistent between reactants and products, including a remapping procedure for ambiguous, potentially mismapped hydrogen atoms. The endpoint geometries can then be optimised and interpolated between, yielding an initial path across the PES for CI-NEB to optimise.}
    \label{fig:ml-kpm-data-overview}
\end{figure}

CRNs generated through Kinetica are stored as xTB-optimised geometries of individual species, along with their SMILES representations; this helps avoid storing every unique system of reactive molecules that is generated. While this is storage-efficient, the required chemical reaction endpoints must be reconstructed from these simpler inputs. An overview of the workflow used for this reconstruction operation is shown in Fig. \ref{fig:ml-kpm-data-overview}.

First, individual species geometries were placed in Cartesian space such that they avoid atomic overlaps and do not spuriously change the character of the reactant species under later geometry optimization. Kinetica includes a simple many-body spring-particle simulator which can be used to optimise the COM of present chemical species while fulfilling this overlap constraint. 

Second, once non-overlapping chemical species for the reactants and products of a given reaction had been created, their atoms were remapped such that they were consistent across a NEB calculation. This was achieved using atom-mapped SMILES, an extension to the regular SMILES representation that uniquely indexes each atom in a structure.\cite{Daylight2011} Kinetica constructs these atom-mapped SMILES when reactions are ingested into a CRN. These can then be used within a substructure search, performed by RDKit, to remap the atom indices of the newly-generated reactant/product systems such that they are consistent across a reaction.\cite{Landrum2024}

However, due to the indistinguishability of hydrogen atoms within regular (unmapped) SMILES and the tendency to treat hydrogen atoms as implicit in many cheminformatics software packages, this approach often leaves hydrogen atoms incorrectly mapped between reactant and product systems. This can result in NEB calculations that erroneously exhibit hydrogen interchange, causing errors in calculated reaction barriers. To remedy this, positions of each pair of hydrogen atoms in the reactant system is permuted. For each permutation, the Kabsch-Umeyama algorithm is used to translate and rotate the modified reactant system in Cartesian space to maximize similarity with the product system.\cite{Kabsch1976,Kabsch1978,Umeyama1991} Here, the hydrogen-index permutation that exhibits the lowest RMSD - as determined using Kabsch algorithm-rotated coordinates - relative to the product structure is then used as the final coordinates (and indices) of the reactants.

The RMSD of atomic positions between the rotated reactant and the product is taken, and if the RMSD of any permutation is smaller than that of the original reactant system and the product system, the positions of the permuted hydrogens are closer to where they should be in a properly atom-mapped reactant/product system pair.

When this occurs, a permutation is made permanent, and this modified reactant system is taken as the base reactant system against which new permutations' RMSD are compared. The hydrogen permutation process continues, iteratively refining the best permutations of hydrogen atoms such that the final reactant system has as low of an RMSD of atomic positions as possible. The Kabsch algorithm-rotated coordinates of this reactant system are taken to be the final coordinates of this endpoint.

Third, after generation of sensible NEB endpoint coordinates, these reactant/product structures were optimised with the NWChem electronic structure code at the same level of theory as the original datasets of Grambow \textit{et al.} against which KPM was fit (namely DFT with $\omega$B97X-D3 hybrid functional and def2-TZVP basis set).\cite{NWChem,Grambow2020,Grambow2020_raddata} This optimization then ensures that the resulting DFT-calculated activation energies are as comparable to the activation energy predictions of KPM as possible.

Finally, after optimization, the reaction endpoints were interpolated using the standard image-dependent pair-potential (IDPP) method. A CI-NEB calculation was then performed for each reaction using the same level of DFT as the endpoint optimizations, with convergence of the calculation controlled by NWChem's \verb|DEFAULT| NEB convergence criteria.

\begin{figure}[p]
    \centering
    \begin{subfigure}[t]{0.49\textwidth}
        \centering
        \includegraphics[width=\textwidth]{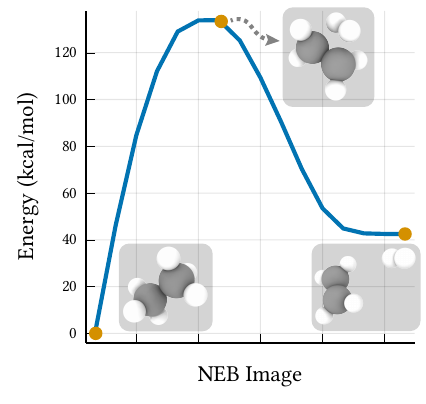}
        \vspace{-20pt}\caption{Single-step concerted dissociation of hydrogen molecule from ethane.}
        \label{fig:ml-kpm-data-rxns-1}
    \end{subfigure}
    \hfill
    \begin{subfigure}[t]{0.49\textwidth}
        \centering
        \includegraphics[width=\textwidth]{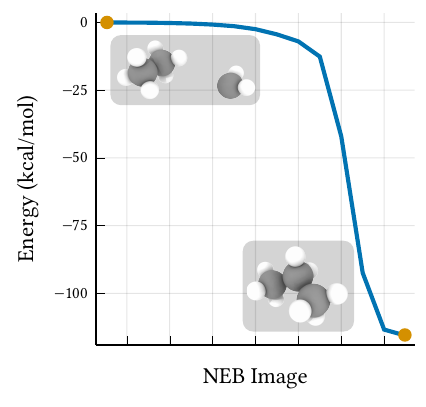}
        \vspace{-20pt}\caption{Barrierless substitution of carbene into ethane C-H bond to form propane.}
        \label{fig:ml-kpm-data-rxns-2}
    \end{subfigure}
    \vskip \baselineskip
    \begin{subfigure}[t]{0.49\textwidth}
        \centering
        \includegraphics[width=\textwidth]{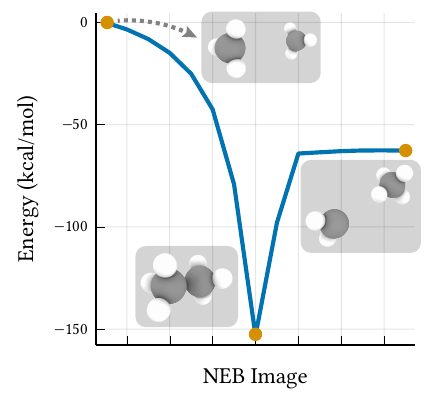}
        \vspace{-20pt}\caption{Multi-step hydrogen abstraction from one methyl radical to another. Proceeds through a stable ethane intermediate.}
        \label{fig:ml-kpm-data-rxns-3}
    \end{subfigure}
    \hfill
    \begin{subfigure}[t]{0.49\textwidth}
        \centering
        \includegraphics[width=\textwidth]{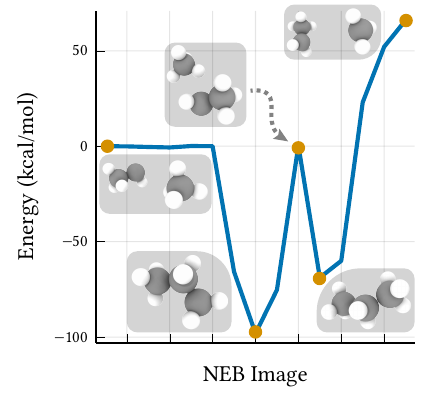}
        \vspace{-20pt}\caption{Multi-step hydrogen abstraction from methane by methylcarbene. Proceeds through two metastable propane intermediates.}
        \label{fig:ml-kpm-data-rxns-4}
    \end{subfigure}
    \caption{A selection of converged CI-NEB energy profiles for chemical reactions generated as part of network $C_4^{I, 0.01}$.}
    \label{fig:ml-kpm-data-rxns}
\end{figure}

A selection of the converged reactions are presented in Fig. \ref{fig:ml-kpm-data-rxns}.

\newpage
\section{KPM Retraining}

\begin{figure}[h!]
    \centering
    \includegraphics[width=240.71031pt]{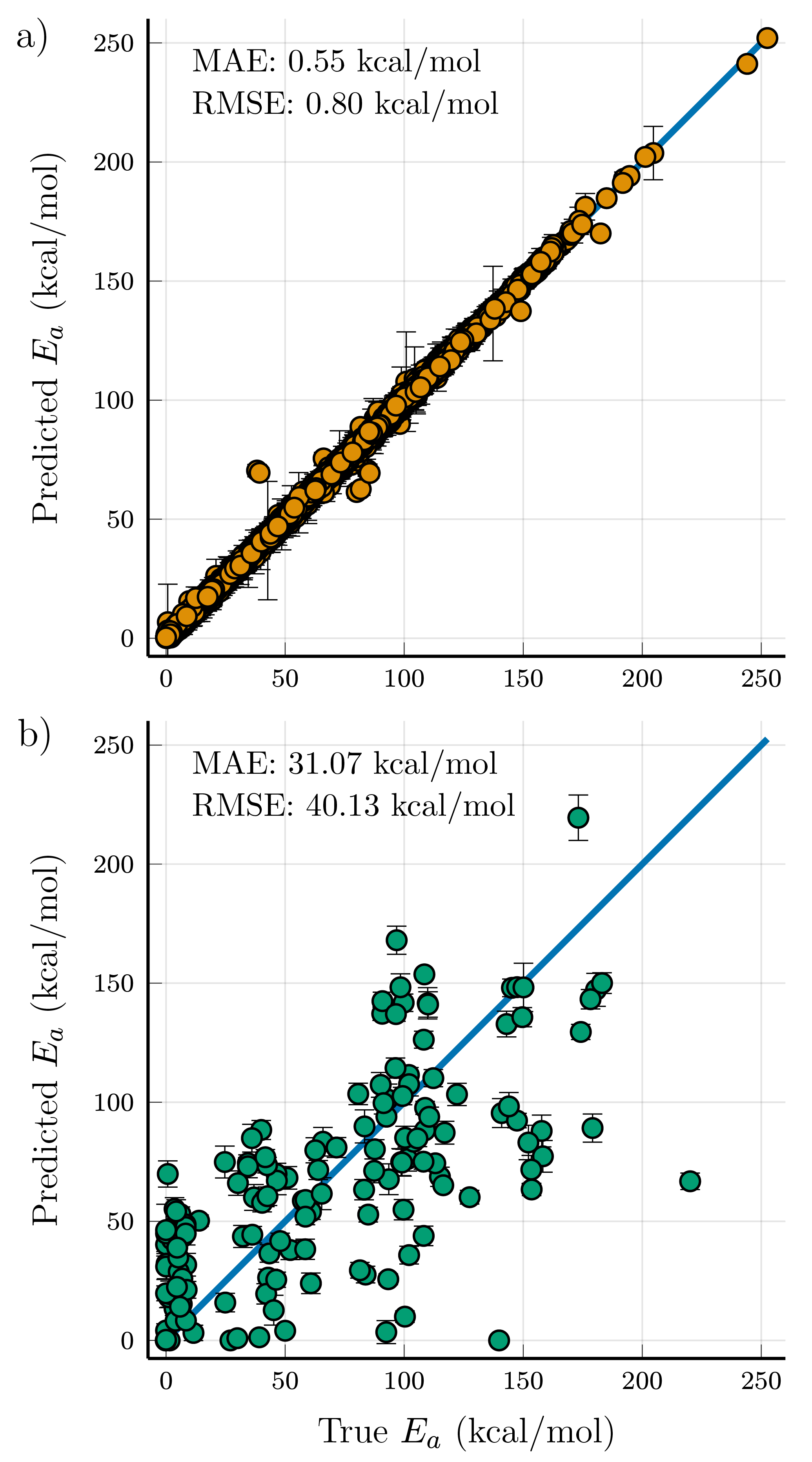}
    \caption{Correlation plots for KPM $E_a$ predictions when retrained with the addition of half of the $C_4^{I, 0.01}$ validation set. a) Predictions for training set. b) Predictions for held-out half of validation set.}
    \label{fig:ml-kpm-ood-retrain}
\end{figure}

The results in Fig. \ref{fig:ml-kpm-ood-retrain} show KPM's predictive performance when its ensemble of 10 NNs was retrained on the radical-extended dataset, plus half of the $C_4^{I, 0.01}$ validation set. 

\newpage
\section{Chemical Reaction Space Analysis}

\begin{figure*}[h!]
    \centering
    \includegraphics[width=0.9\textwidth]{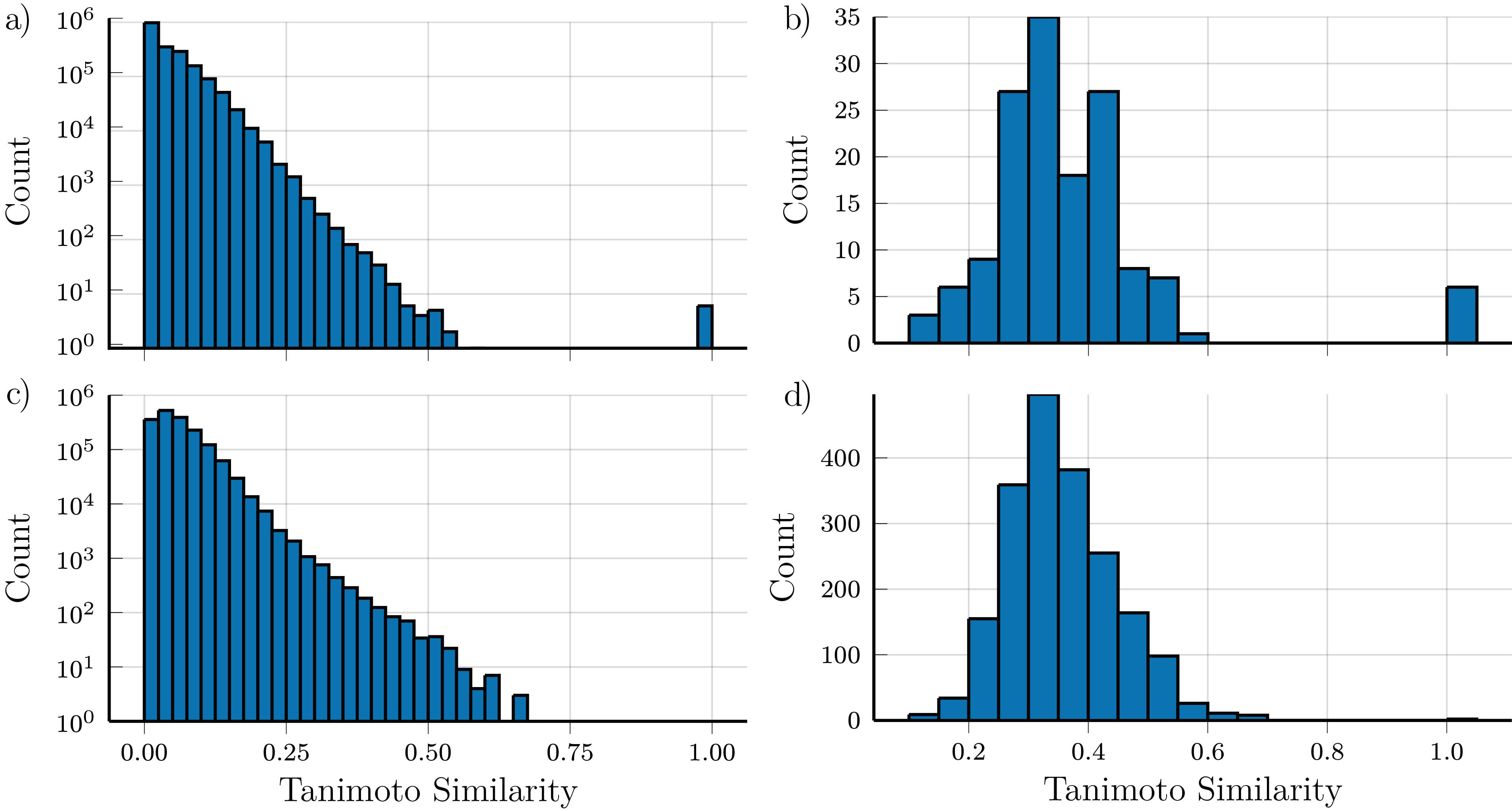}
    \caption{Tanimoto similarity distributions of KPM descriptors in reactions of the radical-extended training set and the $C_4^{I, 0.01}$ validation set. a) Pairwise similarity of all validation set reactions compared to all training set reactions. b) Maximum similarity of validation set reactions to training set reactions. c) Pairwise self-similarity of all training set reactions. d) Maximum self-similarity of training set reactions.}
    \label{fig:ml-kpm-analysis-tanimoto}
\end{figure*}

For vectors of non-negative integers, the Tanimoto coefficient is defined as

\begin{equation}\label{eqn:tanimoto}
    T \left( \mathbf{d}^{\textrm{train}}, \mathbf{d}^{\textrm{val}} \right) = \dfrac{\sum_{i} \textrm{min} \left( d^{\textrm{train}}_i, d^{\textrm{val}}_i \right)}{\sum_{i} \textrm{max} \left( d^{\textrm{train}}_i, d^{\textrm{val}}_i \right)}
\end{equation}

\noindent where $\mathbf{d}^{\textrm{train}}$ is a vector representing the descriptor for a training set reaction, and $\mathbf{d}^{\textrm{val}}$ is a vector representing the descriptor for a validation set reaction. The Tanimoto similarity was calculated for every pair of reactions within and across the two datasets, with results shown in \ref{fig:ml-kpm-analysis-tanimoto}.

Fig. \ref{fig:ml-kpm-analysis-tanimoto}a shows the distribution of pairwise Tanimoto similarities between the reactions of the training set and the reactions of the validation set, while Fig. \ref{fig:ml-kpm-analysis-tanimoto}c shows the self-similarity between all pairs of reactions in the training set. Comparing these distributions does not reveal huge differences; the reactions in the validation set are as similar to the training set reactions as the training set reactions are to themselves. There are some minor differences, and a few validation set reactions are already contained in the training set, indicated by a Tanimoto similarity of 1. However, we do note that the distribution of  Fig. \ref{fig:ml-kpm-analysis-tanimoto}a is slightly skewed towards lower Tanimoto similarity scores, suggesting that some validation-set reactions are quite poorly represented in the training set. Furthermore, Figs. \ref{fig:ml-kpm-analysis-tanimoto}b and \ref{fig:ml-kpm-analysis-tanimoto}d show the \textit{maximum} similarities for every reaction in the validation and training sets, again compared against the reactions in the training set. Again, these two distributions are similar, although the slight shift to lower similarity scores in Fig. \ref{fig:ml-kpm-analysis-tanimoto}b again indicates that the validation-set reactions may exist in a lesser-known area of chemical reaction-space. Overall though, these results show that both sets of distributions are, on the whole, very similar, suggesting that KPM should be able to predict activation energies for the validation set with a similar accuracy to those of the training/test sets --- a trend not observed in the validation set predictions.

\newpage
\section{NeuralNEB performance for selected reactions}

\begin{figure}[h!]
    \centering
    \begin{subfigure}[t]{0.49\textwidth}
        \centering
        \includegraphics[width=\textwidth]{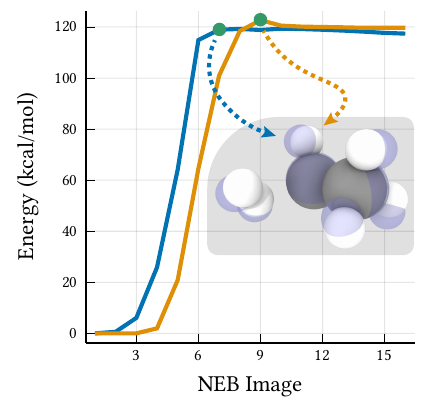}
        \vspace{-20pt}\caption{Molecular hydrogen dissociation from a single carbon atom in ethane to form methylcarbene.}
        \label{fig:ml-bmk-neuralneb_rxns-1}
    \end{subfigure}
    \hfill
    \begin{subfigure}[t]{0.49\textwidth}
        \centering
        \includegraphics[width=\textwidth]{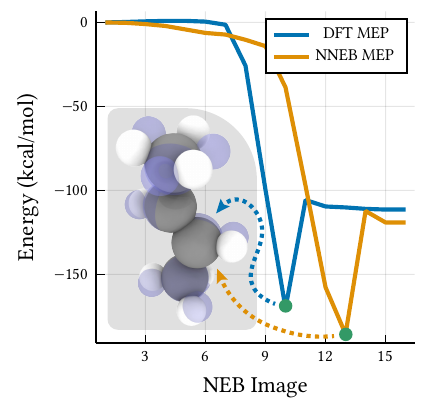}
        \vspace{-20pt}\caption{Insertion of carbon from one methylcarbene into the C-H bond of another. Proceeds through stable intermediate of but-2-ene.}
        \label{fig:ml-bmk-neuralneb_rxns-2}
    \end{subfigure}
    \vskip \baselineskip
    \begin{subfigure}[t]{0.49\textwidth}
        \centering
        \includegraphics[width=\textwidth]{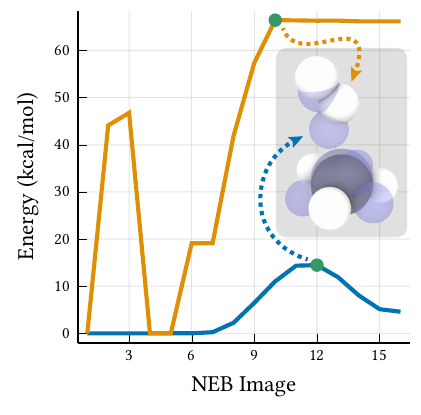}
        \vspace{-20pt}\caption{Abstraction of hydrogen from methane by hydrogen radical, forming methyl radical and molecular hydrogen.}
        \label{fig:ml-bmk-neuralneb_rxns-3}
    \end{subfigure}
    \hfill
    \begin{subfigure}[t]{0.49\textwidth}
        \centering
        \includegraphics[width=\textwidth]{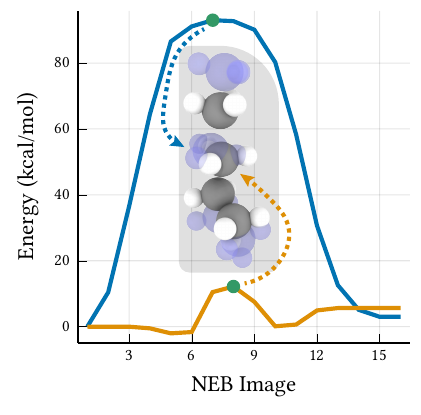}
        \vspace{-20pt}\caption{Abstraction of methyl group from ethane by ethyl radical, forming propane and methyl radical.}
        \label{fig:ml-bmk-neuralneb_rxns-4}
    \end{subfigure}
    \caption{A selection of converged DFT and NeuralNEB (shortened here to NNEB) CI-NEB energy profiles for chemical reactions which highlight some of the similarities and differences between the two ESMs. Where TS geometries have been overlaid, the translucent blue `ghost' atoms represent the DFT TS.}
    \label{fig:ml-bmk-neuralneb_rxns}
\end{figure}

To explore why NeuralNEB fails to capture reaction energetics to such a degree, individual MEPs were analysed to determine the where it was misrepresenting the PES in comparison to DFT. Examples of common areas of difficulty are shown in \ref{fig:ml-bmk-neuralneb_rxns}. In each case, the energies of the optimised reactants have been aligned and the TSs located by DFT and NeuralNEB have been highlighted and visualised. In cases where there is clear overlap, the TSs have been overlaid to visually represent the differences in the PESs that the two methods are operating on; where this occurs, the DFT TS is shown as a translucent blue set of `ghost' atoms, while the NeuralNEB TS is opaque.

Figure \ref{fig:ml-bmk-neuralneb_rxns-1} depicts a molecule of hydrogen breaking off from a single methyl carbon in ethane to form methylcarbene. Both the MEPs generated by the two methods and their respective TSs show a strong agreement, with only a $3.83$ kcal/mol deviation in activation energy. Many of the MEPs converged by NeuralNEB exhibit this strong agreement, although these all tend to share the same reaction energy profile and are either barrierless, or the reverse of a barrierless reaction (as depicted here). This implies that NeuralNEB has a strong understanding of barrierless, single-step reactions.

Figure \ref{fig:ml-bmk-neuralneb_rxns-2} shows a multi-step reaction, whereby two methylcarbene molecules come together in an insertion reaction. The carbene carbon on one molecule inserts into the carbene C-H bond of the other, ultimately forming ethylmethylcarbene; however, this proceeds through a stable but-2-ene intermediate. At first glance, NeuralNEB captures both the energetic and geometric profiles of this reaction remarkably well for a MLIP, with the geometry of the intermediate almost exactly matching except for a rotated methyl end group. However, the quantitative difference between the reaction energetics is quite high --- the intermediates differ by $15.45$ kcal/mol --- owing to the large energy space covered.

In Fig. \ref{fig:ml-bmk-neuralneb_rxns-3}, a hydrogen atom on methane is abstracted by a hydrogen radical. Despite this being one of the simplest radical reactions one could envision, NeuralNEB completely fails to accurately predict this MEP. Instead, it initially strongly overbinds the radical hydrogen to the carbon, before the departing hydrogen is removed alongside the original radical with a very high energy compared to DFT. This dissociated state then remains high in energy while the separation between the two molecules increases, while in the DFT MEP it falls as they separate into molecular hydrogen and a methyl radical. This propensity to incorrectly evaluate the energetics of species with single radical electrons is repeated in other reactions within the converged set, and indicates that such monoradicals are not present in the Transition1x dataset. 

Finally, Fig. \ref{fig:ml-bmk-neuralneb_rxns-4} shows a methyl group abstraction from ethane, performed by an ethyl radical, forming propane and a methyl radical. The DFT TS shows the abstracted methyl group lying between the pseudo-radical ethyl and methyl moieties, which notably never closely approach one another. By contrast, the NeuralNEB TS is created by this close approach, with the reactants coming together to form an over-protonated analogue of butane before separating again. As with the previous reaction, this also involves both radical reactants and radical products, and NeuralNEB completely fails to capture the reaction energetics, further confirming that any energies of such geometries must be extrapolative and not trained upon.

\newpage
\section{Reaction Interpolation}

In addition to the $r_{\textrm{SOAP}}$ metric described in the main text, we also compared TS geometries with the RMSD between the positions of the atoms in the two geometries (which had their translational and rotational differences minimized with the Kabsch-Umeyama algorithm):

\begin{equation}\label{eqn:ml-spahm-pos_rmsd}
    r_{\text{RMSD}} = \frac{1}{n}\sum_{i=1}^{n}\left \| \mathbf{r}_{i}^{\textrm{TS}} - \mathbf{r}_{i}^{\textrm{interp}} \right \|^2
\end{equation}

\noindent $n$ is the number of atoms in a given TS and $\mathbf{r}_{i}^{\textrm{TS}}$ and $\mathbf{r}_{i}^{\textrm{interp}}$ are vectors of atomic positions for the DFT TS and the interpolated TS, respectively. The resulting value $r_{\text{RMSD}}$ is interpretable as the average distance that the atoms in the interpolated TS lie away from the atoms in the true TS in Cartesian space.

\begin{table}[h]
    \centering
    \begin{tabular}{@{}lrr@{}}
        \toprule
        Interpolation Type & $r_{\text{RMSD}}$ ($\textnormal{\r{A}}$)  & $r_{\textrm{SOAP}}$ \\
        \midrule
        Linear & $0.51 \pm 0.27$ & $42.51 \pm 61.45$ \\
        IDPP & $0.55 \pm 0.33$ & $53.10 \pm 48.41$ \\
        Geodesic & $0.44 \pm 0.26$ & $27.61 \pm 21.83$ \\
        \bottomrule
    \end{tabular}
    \caption{Statistics for geometric difference metrics between true and interpolated TSs.}
    \label{tab:ml-spahm-interp-dists}
\end{table}

\begin{table}[p]
    \centering
    \includegraphics{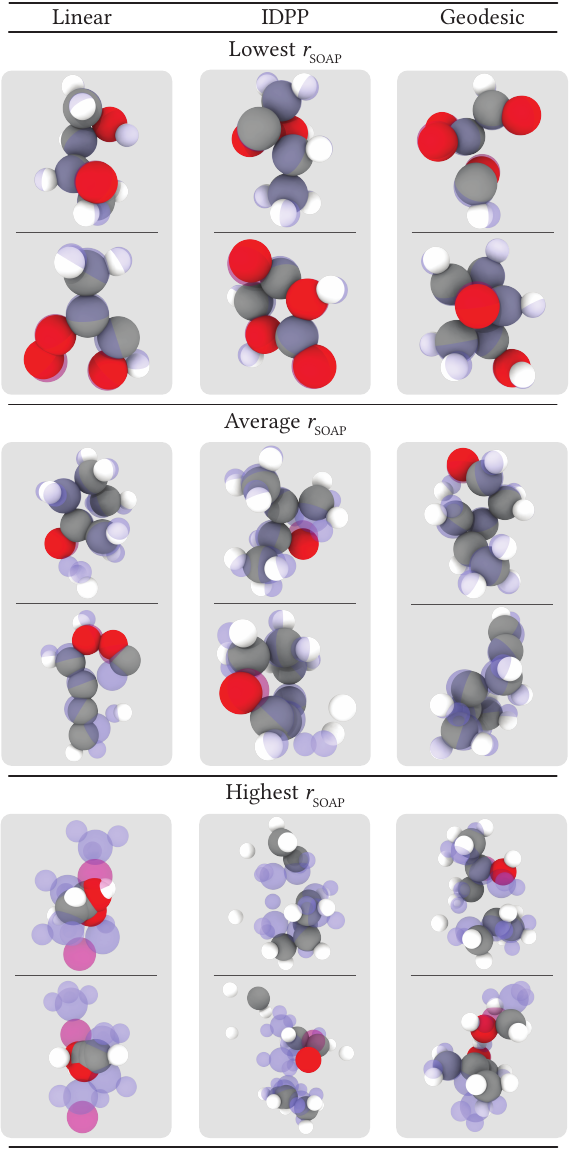}
    \vspace{-5pt}\caption{Examples of TS approximations generated by each interpolation method (opaque), overlaid with their DFT TS geometries (translucent), for minimum, average and maximum $r_{\textrm{SOAP}}$.} 
    \label{tab:ml-spahm-interp-examples}
\end{table}

For the case of the approximate TSs with the lowest values of $r_{\textrm{SOAP}}$, all interpolation methods tested estimated TS geometries with excellent accuracy. The only observable differences in these geometries were small deviations in hydrogen atom placements, which were present across all methods.

For the TSs with $r_{\textrm{SOAP}}$ closest to their interpolation method's average, the differences become more apparent. Minor hydrogen misplacement is still present across all methods, but more important is the difference in atomic connectivity also observable across the board. This ranges from major changes, such as missing C-O bonds in the case of the linear interpolations and missing C-H bonds in IDPP, to more minor changes, such as a hydrogen atom being more weakly bound to an oxygen atom than expected in one of the geodesic interpolations. All the interpolation methods also experience conformational differences of varying degrees.

At the high end of $r_{\textrm{SOAP}}$, the differences between the methods become quite stark. The linear interpolations in this regime are all the result of atomic overlap, a common consequence of this method in reactions where there is a significant geometric change between reactants and products as interpolated atoms move in a straight line between these states, irrespective of other atoms in the way. Meanwhile, IDPP experiences extreme distortions in its estimated TSs, with hydrogen atoms scattered around the geometries and frequently disconnected carbon and oxygen atoms. This is an unfortunately common occurrence when using IDPP for complex reactions, where oscillations can build up over the course of an optimisation that ultimately result in a rapid explosion of atomic forces and therefore seemingly random, fast movements of atoms and a failure to converge to a stable MEP.\cite{Zhu2019} Avoiding this problem is one of the targets of the geodesic interpolation method, and it is able to entirely avoid completely non-physical geometries as a result. Instead, geodesic interpolation's worst deviations from true TS geometries are mostly conformational, and while these would undoubtedly lead to calculation of significantly different activation energies under DFT, this calculation would at least be possible.

\newpage
\section{DCINet Training}

\begin{figure}[h]
    \centering
    \includegraphics{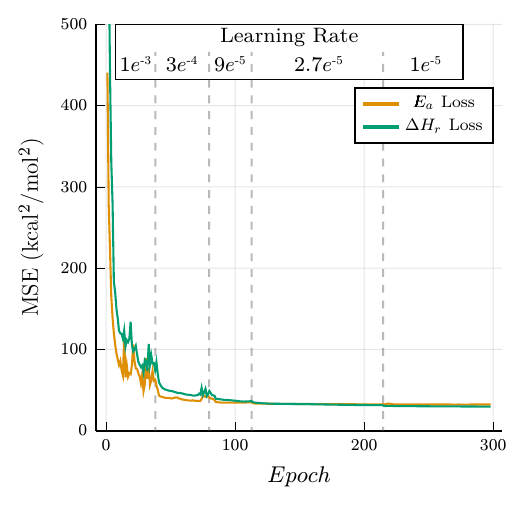}
    \caption{Loss during training for the loss components of the best DCINet model, with changes in learning rate annotated.}
    \label{fig:dcinet-lr}
\end{figure}

\begin{figure*}[h]
    \centering
    \includegraphics[width=0.9\textwidth]{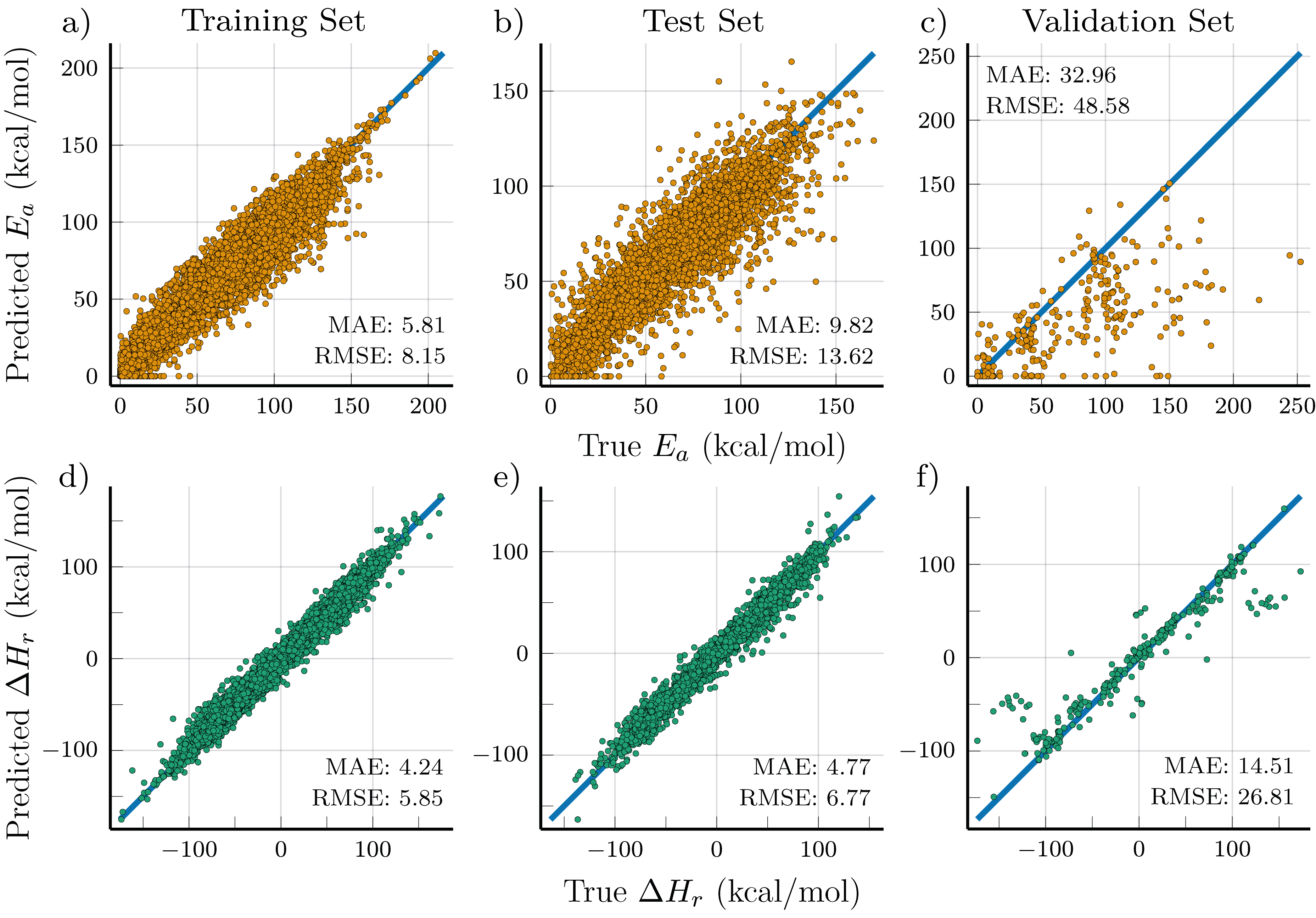}
    \caption{Correlation plots for a)--c) $E_a$ and d)--f) $\Delta H_r$ predictions of a variant DCINet model without TS information, when applied to the training set, test set and $C_4^{I, 0.01}$ validation set respectively. All error metrics are given in units of kcal/mol.}
    \label{fig:ml-spahm-nots-conv-corr}
\end{figure*}

\newpage
\bibliographystyle{achemso}
\bibliography{bibliography}

\end{document}